\documentclass[pdflatex, sn-basic]{sn-jnl}

\usepackage{setspace}
\usepackage{amsmath,amsfonts,amssymb,physics,dirtytalk}%
\usepackage{graphicx}%
\usepackage{subfigure}
\usepackage{subfloat,siunitx}
\usepackage{float}
\usepackage{setspace}
\usepackage{hyperref}
\usepackage{cleveref}
\usepackage{xcolor}%
\usepackage{multirow}%
\usepackage{comment}
\usepackage{amsthm}%
\usepackage{mathrsfs}%
\usepackage[title]{appendix}%
\usepackage{textcomp}%
\usepackage{manyfoot}%
\usepackage{booktabs}%
\usepackage{algorithm}%
\usepackage{algorithmicx}%
\usepackage{algpseudocode}%
\usepackage{listings}%

\newcommand{\fzu}[1]{$\ \si{erg\ cm^{-2}\ s^{-1}}${#1}} 

\newcommand{\amdl}{\textbf{GDS-140}}

\newcommand{\mgds}{\textbf{GDS-140}$^{\boldsymbol{1}}$}
\newcommand{\mgdsthick}{\textbf{GDS-140}$^{\boldsymbol{1.5}}$}
\newcommand{\mstack}{\textbf{GDS-140}$\boldsymbol{\ell_2^1}$}
\newcommand{\mstackthick}{\textbf{GDS-140}$\boldsymbol{\ell_2^{1.5}}$}
\newcommand{\mstackthree}{\textbf{GDS-140}$\boldsymbol{\ell_3^{1}}$}

\begin{document}
\title[CZT]{Evaluation of the performance of a CdZnTe-based soft $\gamma$-ray detector for CubeSat payloads.}

\author[1]{\fnm{Kees} \sur{de Kuijper}} \email{kdkuijper@gmail.com}
\author[1]{\fnm{Rishank} \sur{Diwan}} \email{rishank.phy.iitkgp@gmail.com}
\author*[1,3]{\fnm{Partha Sarathi} \sur{Pal}} \email{parthasarathi.pal@gmail.com}
\author[1]{\fnm{Andreas} \sur{Ritter}}
\author[2]{\fnm{Pablo M.} \sur{Saz Parkinson}}
\author[1]{\fnm{Andy~C.~T.} \sur{Kong}}
\author[1]{\fnm{Quentin~A.} \sur{Parker}}

\affil*[1]{\orgdiv{Laboratory for Space Research, Faculty of Science,}, \orgname{The University of Hong Kong}, \orgaddress{\street{405B, Block A, Cyberport 4, 100 Cyberport Road}, \city{Hong Kong}, \country{China}}}

\affil[2]{\orgdiv{Santa Cruz Institute for Particle Physics and Department of Physics}, \orgname{University of California}, \orgaddress{\street{Street}, \city{Santa Cruz}, \postcode{95064}, \state{CA}, \country{USA}}}

\affil[3]{\orgname{Institute of Astronomy Space and Earth Science}, \orgaddress{\street{P 177, CIT Road, Scheme 7M}, \city{Kolkata}, \postcode{700054}, \state{West Bengal}, \country{India}}}

\abstract{The low-energy $\gamma$-ray (0.1-30 MeV) sky has been relatively unexplored since the decommissioning of the COMPTEL instrument on the Compton Gamma-Ray Observatory (CGRO) satellite in 2000. However, the study of this part of the energy spectrum (the \say{MeV gap}) is crucial for addressing numerous unresolved questions in high-energy and multi-messenger astrophysics. Although several large MeV $\gamma$-ray missions like AMEGO and e-ASTROGAM are being proposed, they are predominantly in the developmental phase, with launches not anticipated until the next decade at the earliest. In recent times, there has been a surge in proposed CubeSat missions as cost-effective and rapidly implementable ``pathfinder" alternatives. A MeV CubeSat dedicated to $\gamma$-ray astronomy has the potential to serve as a demonstrator for future, larger-scale MeV payloads. This paper presents a $\gamma$-ray payload design featuring a CdZnTe crystal calorimeter module developed by IDEAS. We report the detailed results of simulations to assess the performance of this proposed payload and compare it with those of previous $\gamma$-ray instruments.}


\maketitle

\section{Introduction}
\label{sect:intro}
In the past two decades, the field of $\gamma$-ray astronomy has experienced significant development, leading to the exploration of the high-energy Universe with unprecedented sensitivity. Since its launch in 2008, the \textit{Fermi} Large Area Telescope (LAT)~\citep{Atwood09} has been continuously monitoring the entire $\gamma$-ray sky every three hours, covering the energy range from tens of MeV to $>$300 GeV. Throughout this period, the LAT has detected $\gamma$ rays from a multitude of source categories, including pulsars~\citep{smith2023fermi}, Blazars~\citep{Kerby_2023, 2014ApJ...780...73A} and other Active Galactic Nuclei (AGN)~\citep{Ajello_2020}, Gamma-ray Bursts (GRB)~\citep{Ajello_2019}, and even our own Galactic Center~\citep{Ackermann_2017, PhysRevD.103.063029}.
\textit{Fermi} observations have also uncovered a plethora of unforeseen discoveries, including the \textit{Fermi} Bubbles~\citep{galaxies6010029, Su_2010, ackermann2014spectrum}, a large population of radio-quiet $\gamma$-ray pulsars~\citep{Abdo09,SazParkinson10}, classical $\gamma$-ray novae~\citep{Ackermann14}, a multitude of extreme blazars/AGN~\citep{Ackermann_2015}, high-energy solar flares~\citep{Ajello_2021}, $\gamma$-ray flashes in thunderstorms~\citep{https://doi.org/10.1029/2017JA024837}, etc. Furthermore, \textit{Fermi} observations have been critical to many recent groundbreaking advancements in the burgeoning field of multi-messenger astronomy~\citep{Abbott17,TXS0506, PhysRevD.103.123005}.


Despite all these remarkable achievements in {\it high-energy} ($>$30 MeV) $\gamma$-ray astronomy, the {\it soft} $\gamma$-ray sky (0.1 to 30 MeV) has, by contrast, been relatively unobserved. Since the de-orbiting of the imaging Compton telescope (COMPTEL)~\citep{Schoenfelder93} on board the Compton Gamma-ray Observatory (CGRO), in 2000, there have been no follow-up missions with improved sensitivity, resulting in an observational gap in this energy range, often referred to as the \say{MeV gap}. 

It is of paramount importance, more than two decades later, to conduct a comprehensive exploration of this energy range to gain a better understanding of astrophysical sources and the rapidly-growing field of {\it multi-messenger} astronomy. The investigation of MeV pulsars, blazars, GRBs, and the MeV background are all major research topics for the $\gamma$-ray community. Finally, the study of the 511 keV emission line from electron-positron annihilation would also allow us to gain a better understanding of systems such as SNIa\citep{2018MNRAS.480.1393C} and Galactic compact objects\citep{2011A&A...531A..56D}. 

Despite the large number of $\gamma$-ray pulsars detected by the LAT, a number of crucial questions remain, including the location and main production mechanism of the $\gamma$-ray emission, as well how such a mechanism the evolution of rotation-powered pulsars with age. Many of these questions might be answered through the study of a large population of MeV pulsars that may be present in our Galaxy, but beyond the sensitivity of \textit{Fermi}~\citep{Astro2020}.

Blazars represent the most numerous class of high-energy $\gamma$-ray sources. Observations of blazars in the MeV gap will serve to link X-ray and $\gamma$-ray portions of the Spectral Energy Distributions (SEDs) and help explain the Blazar SED evolution~\citep{2017MNRAS.469..255G}. Numerous Blazar SEDs are anticipated to reach maximum values around a few MeV, a region currently unobservable~\citep{Ajello_2016, Marcotulli_2020}. The inclusion of MeV photons in blazar SED modeling is also of great significance given the that blazars are plausible neutrino sources~\citep{TXS0506}. Thus, Blazars are prime targets in the era of multi-messenger astronomy~\citep{boettcher2022multiwavelength}. It should be noted that neutrino observations require simultaneous $\gamma$-ray observations to effectively search for counterparts. To date, however, there are no definitive correlations between neutrino observations and $\gamma$-ray photons, albeit tentative associations have been reported~\citep{TXS0506}. Thus, a counterpart search in the MeV energy band is crucial to better understand future neutrino events~\citep{murase2016hidden}.

MeV observations also provide the opportunity to explore our Galactic neighborhood. Studies of the 1.8 MeV Galactic Aluminum-26 emission line, for example, can be used to trace the nucleosynthesis of radioactive elements in the vicinity of the Galactic Center~\citep{Beechert_2022}. 

The elusive MeV background is another topic that is ripe for inquiry. Although it is thought to be composed of MeV photons originating from a collection of unresolved sources, such as Cosmic rays, AGN, SNIa, star-forming galaxies, among others~\citep{Ruiz-Lapuente_2016}, it is only with fresh observations that the present models can be better constrained and other alternatives ruled out.

A number of major MeV $\gamma$-ray missions are in various stages of development. These include AMEGO~\citep{AMEGOref}, and e-ASTROGAM~\citep{deAngelis18}. Unfortunately, it is unlikely that any of these will be launched in the coming decade. At a smaller scale, the Compton Spectrometer and Imager (COSI) mission~\citep{tomsick2019compton, Tomsick:2021H5} is a small explorer mission, currently under development by NASA, which will explore the 0.2-5 MeV energy range.

Several CubeSat configurations are also undergoing development at varying stages~\citep{Racusin17, Wen21, Bloser2022}. Among the projects is a Double-Sided Silicon Strip Detector (DSSD) payload design, which has the added capacity to detect polarization from $\gamma$-ray sources within the MeV range~\citep{Yang_2020, diwan20236u}.
Another alternative $\gamma$-ray detector technology within CubeSat dimensions with a CdZnTe crystal is also proposed \citep{2022JCAP...08..013L, 2022JInst..17P8004L}. 
Presently, the detector design is improved, and a better CdZnTe module design is proposed by IDEAS\footnotemark[1] \footnotetext[1]{\url{https://ideas.no/products/gds-100/}} in collaboration with Redlen Technologies\footnotemark[2] \footnotetext[2]{\url{https://redlen.com/}}. This new design is also in consideration for future space missions \citep{mass2023}. In this paper, the design and performance of a CdZnTe crystal module, based on the proposed latest design from IDEAS, for a CubeSat mission, which can perform observations within the desired energy range is presented. This CubeSat mission could serve as a pathfinder to estimate the performance of larger-scale MeV telescopes.

\section{Detector Design}
A Compton telescope is an instrument specifically designed to detect MeV $\gamma$-rays by utilising Compton scattering within high atomic number (Z) detectors. At present, Cadmium Zinc Telluride (CdZnTe, hereafter CZT) semiconductor detectors have become increasingly popular due to their high detection efficiency, room-temperature operation, wide field-of-view, and favourable angular resolution ($\mathcal{O}$($1^{\circ}$)), making them well-suited for various X-ray and $\gamma$-ray detection applications \citep{2010NIMPA.623..434A, SCHLESINGER2001103, 2009Senso...9.3491D}. The high atomic number and density of CZT, when compared to traditional semiconductor materials such as silicon (Si) and germanium (Ge), make it a highly advantageous choice for radiation detection. Moreover, CZT detectors can be produced with exceptional quality and can range up to a few cubic centimeters in volume using current technology.

The proposed payload in this paper is intended to satisfy the requirements of a CubeSat mission, which is a class of nano-satellites possessing a standardised size and form factor. The standard size for a CubeSat is one \say{unit} (1U) and possesses a volume of $10 \times 10 \times 10\ \si{cm^3}$ 
and maximum weight of 1.33 kg. Additionally, it is feasible to amalgamate multiple units, with the current CubeSat Design Specification\footnotemark[3] \footnotetext[3]{\url{https://www.cubesat.org/cubesatinfo}} defining envelopes for 1U, 1.5U, 2U, 3U, and 6U form factors, and the possibility of extensions up to 12U and 16U. CZT detectors are highly suitable for the creation of compact and reliable radiation detection systems, with a broad range of applications that include nuclear medicine, radiation monitoring, and X-ray and $\gamma$-ray astronomy.

\begin{figure*}[h]
\centering 
\includegraphics[trim={1.5cm 3.5cm 1.5cm 4cm},clip,width=0.9\textwidth]{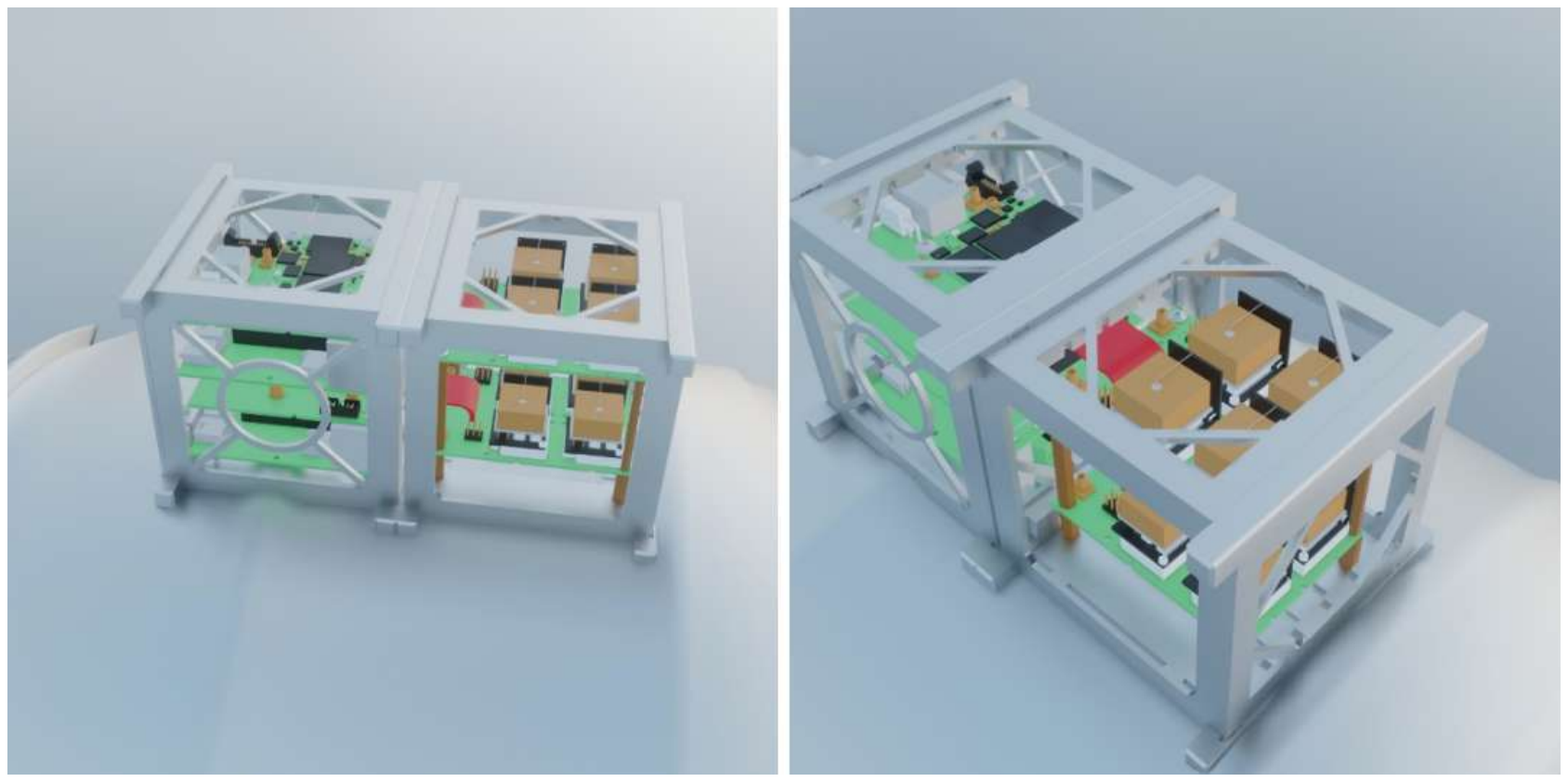}
\caption{Rendered diagram of the assembled payload inside a 2U Cubesat casing with two \amdl\ modules (shown in the right half of the 2U casing). The CZT crystals are contained within the brown ``boxes". Out of the whole 2U volume, 1U is dedicated to housing the \amdl\ modules and another 1U is dedicated to the HV power supply unit.}
\label{rendered_design}
\end{figure*}

In the proposed payload design, an improved crystal specification has been used, which is better than its predecessor~\cite{2022JCAP...08..013L, 2022JInst..17P8004L}. The upgraded CZT detector provided by Redlen has a volume of $2.0 \times 2.0 \times 1 (1.5)\ \si{cm^3}$  (P/N R14892 (P/N R14892$\_$15)), employing an $11 \times 11$ pixel anode structure and 2 cathode channels. The pixel pitch is 1.72 mm and the pixel size is 1.22 mm. For each event interaction, the triggered pixels provide information on the deposited energy and location on the anode plane (x-y plane), while the interaction depth (location on the z-axis) is reconstructed from the ratio between the cathode signal and pixel signal. Pixels from each CZT crystal detector are read out by the 121-channel GDS-10 developed by IDEAS, on the flip side of the motherboard, GDS-100\footnotemark[1], hosting the detector itself. Each GDS-100 motherboard can hold up to 4 CZT crystals mounted on GDS-10, with a pitch of 32 mm. This configuration of 4 CZT crystals mounted on a GDS-100 motherboard is referred to as \amdl\ in this paper. The controller board and high-voltage components are provided on a second board at the side of the payload. Figure~\ref{rendered_design} contains the 3D rendered image of the CubeSat payload proposed to be launched. The 3D rendered design includes the two-layer \amdl\ module and the electronics included.



The Compton detector is enclosed from the top and 4-sides by an anti-coincidence detector (ACD), which is essential to veto the in-orbit cosmic-ray background. The technology is composed of several slabs of plastic scintillators, each 5 mm thick, read out by silicon photo-multipliers and has been implemented in other previous missions, such as \textit{Fermi}-LAT~\citep{Atwood09} and AGILE~\citep{2009A&A...502..995T}. The rejection efficiency is expected to be $\gtrsim 99.99\%$. Principally, for these detector designs, only active materials have been simulated, ignoring structural elements and the fact that some space will be reserved for the read-out electronics and payload setup. Based on these assumptions, it is anticipated that the former will have no impact on the simulation and can be neglected. The compatible design of the CubeSat allows for elementary scaling into different payload sizes. For instance, various geometries with different units can be utilised for different missions with different requirements.


\begin{table}[ht]
\caption{Simulated payload designs with different \amdl\ configurations shown in Figure~\ref{payload_design}.} 
\label{design_table}
\begin{tabular*}{\textwidth}{@{\extracolsep\fill}cccccccc} 
\toprule
Payload name & Design & Det. & Crystal & CZT & Millipore & NE110 & Total\\ 
 & (\#) & Vol & Size ($\si{cm^3}$) & (g) & (g) & (g) & (g) \\ \midrule
\mgds & 1 & 1U & $2\times2\times1$ & 97.0 & 21.2 & 93.9 & 212.0 \\ 
\mgdsthick & 2 & 1U & $2\times2\times1.5$ & 145.4 & 21.2 & 93.9 & 260.5 \\ 
\mstack & 3 & 1U & $2\times2\times1$ & 193.9 & 42.4 & 190.6 & 426.9 \\ 
\mstackthick & 4 & 1U & $2\times2\times1.5$ & 290.9 & 42.4 & 190.6 & 523.8 \\
\mstackthree & 5 & 1U & $2\times2\times1$ & 290.9 & 63.6 & 190.6 & 545.0 \\
\botrule
\end{tabular*}
\end{table}

\subsection{Different payload designs}
\begin{figure*}
\begin{minipage}[c][0.9\textheight]{\textwidth}
\centering
\subfigure[\mgds]{\label{gds_100}
\includegraphics
[trim={5cm 1.5cm 4.5cm 3.2cm}, clip, 
scale=0.45]{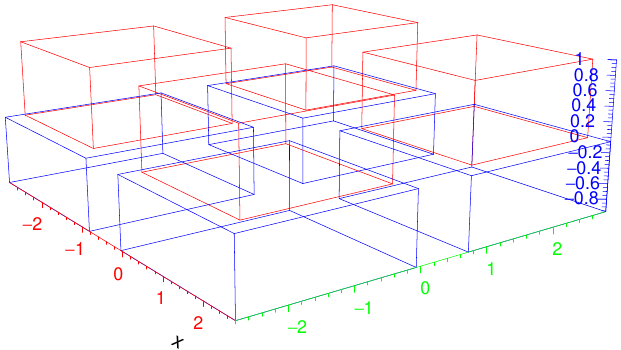}}
\subfigure[Reconstructed point source \mgds]{\label{gds_100_psf}
\includegraphics[trim={0.7cm 0.5cm 0cm 2.5cm}, clip, width=0.425\textwidth]{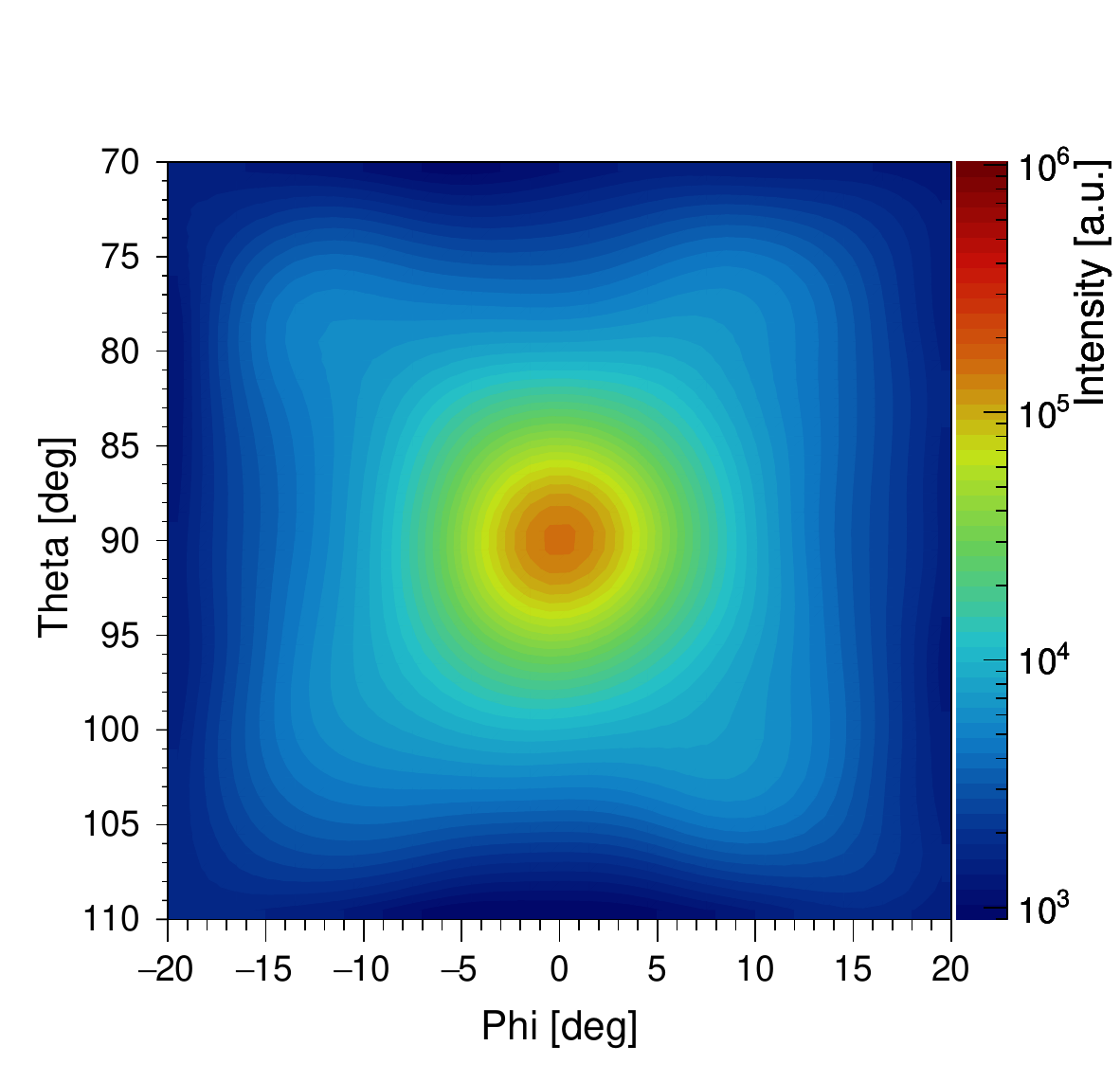}}
\subfigure[\mstack]{\label{stack_2}
\includegraphics
[trim={6.5cm 1cm 6.5cm 2cm}, clip, 
scale=0.65]{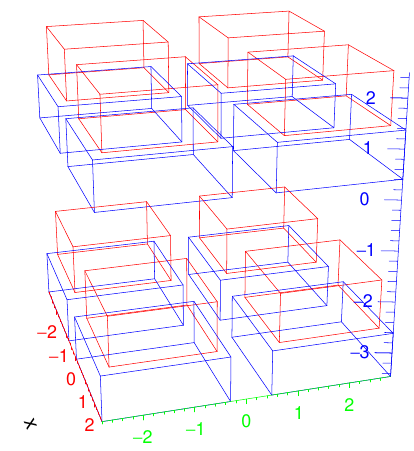}}
\subfigure[Reconstructed point source \mstack]{\label{stack_2_psf}
\includegraphics
[trim={0.7cm 0.5cm 0cm 2.5cm}, clip, width=0.425\textwidth]{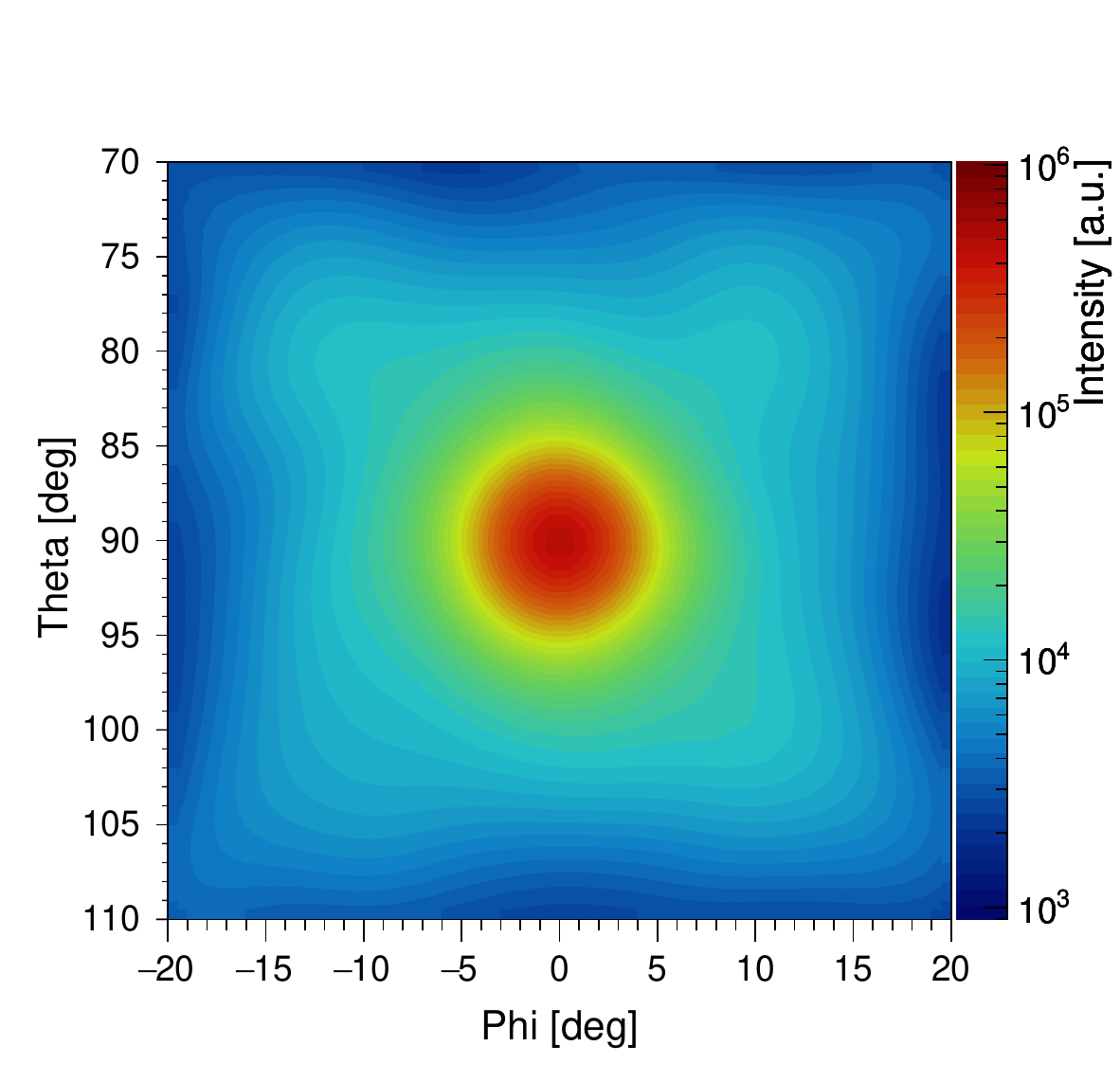}}
\subfigure[\mstackthree]{\label{stack_3}
\includegraphics
[trim={6.8cm 1cm 7cm 2cm}, clip, 
scale=0.65]{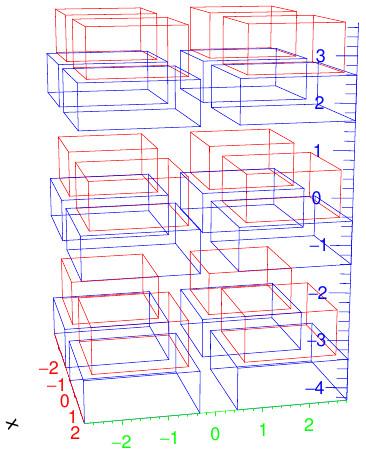}}
\subfigure[Reconstructed point source \mstackthree]{\label{stack_3_psf}
	\includegraphics[trim={0.7cm 0.5cm 0cm 2.5cm}, clip, width=0.425\textwidth]{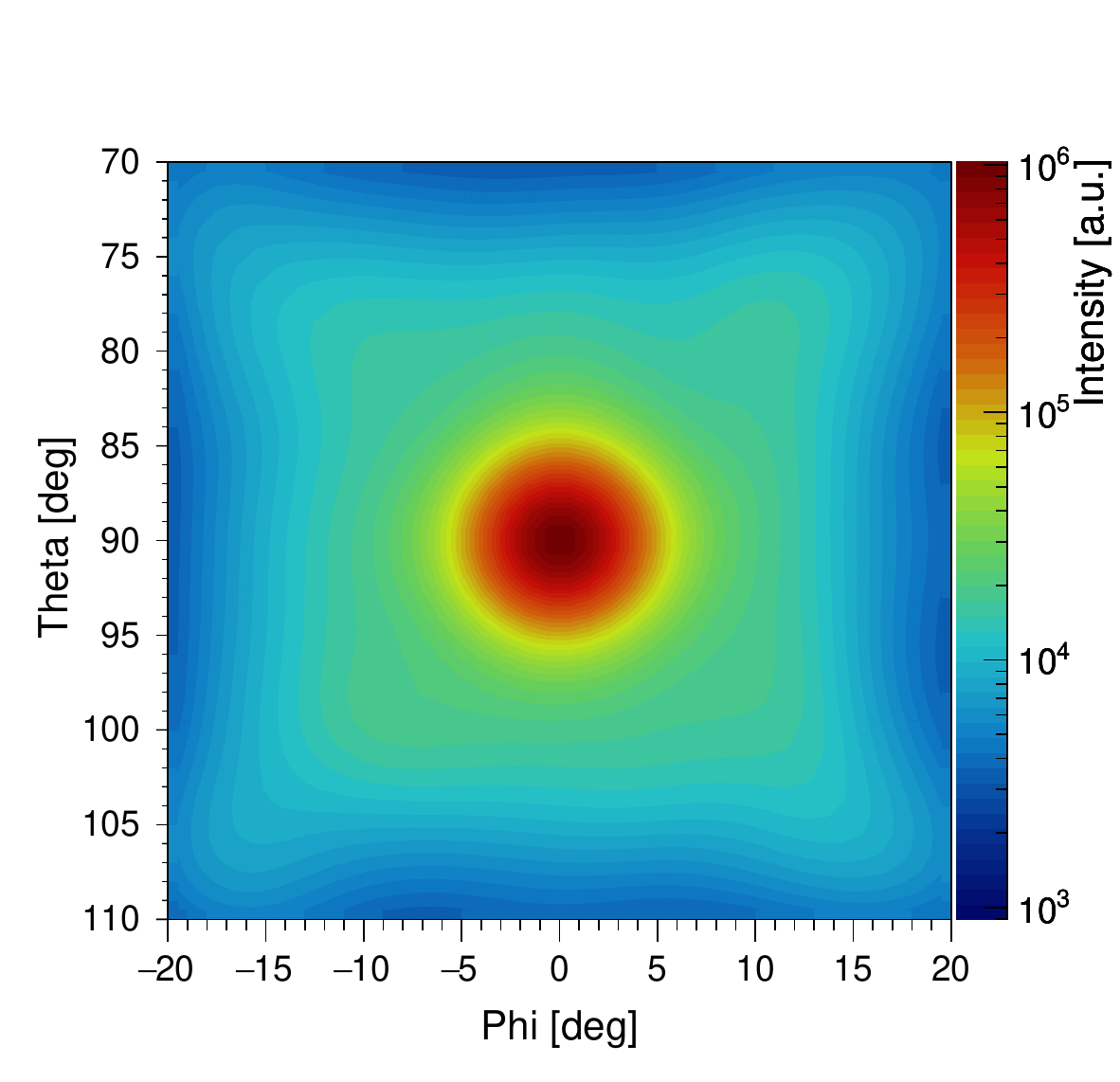}}

\caption{Illustration of different \amdl\ payload designs with simulated point source image reconstruction performance. The left column shows the different payload designs with crystal (shown in red) thickness of 1 cm.
The blue block represents the CZT crystal mount and readout (GDS-10). 
The scales along the axes are in cm.
The right column shows the simulated reconstructed far-field point source image from the payload designs. To better compare and distinguish source intensity differences for the three payload designs, the minimum and maximum intensity was kept the same and logarithmic scaling was used. A clear improvement in source intensity can be noticed with increasing number of crystal layers.
\label{payload_design}}
\end{minipage}
\end{figure*}

Different payload designs are useful in selecting the optimum design for a future mission. Moreover, this paper discusses five different designs whose specifications are shown in Table~\ref{design_table}. Three of those designs (\#1,3,5) are shown on the left column of Figure~\ref{payload_design}. Of these five different designs, the first two designs (\#1-2) are the standard designs, and the other three (\#3-5) are \say{extended} designs. All of these \amdl\ designs fit into a 1U volume. 
Designs \#1-2 incorporate a single \amdl, one with crystal depth $d=1\ \si{cm}$ (denoted \mgds, shown in Figure \ref{gds_100}) 
and another with $d=1.5\ \si{cm}$ (denoted \mgdsthick). Both of these designs use the same GDS-100 chip and electronics. The design for \mgdsthick\ is the same as in Figure \ref{gds_100}, except the crystal depth is 1.5 cm instead of 1 cm.
Designs \#3 and \#4 incorporate 2 layers ($\ell$) of \amdl, again both using a crystal depth $d=1\ \si{cm}$ and $d=1.5\ \si{cm}$, and are denoted as \mstack\ (shown in Figure \ref{stack_2}) and \mstackthick\ respectively. In the \mstackthick\ design, the $d=1$ cm crystals in Figure \ref{stack_2} are replaced with $d=1.5$ cm crystals. 
The fifth and last design uses 3 layers of \amdl, shown in 
Figure \ref{stack_3}, and is denoted here as \mstackthree. In order to fit this design into a 1U volume, a crystal depth of $d=1\ \si{cm}$ is adopted. On the right column of Figure \ref{payload_design}, the simulated reconstructed intensity distribution of a far-field point source for payload designs \#1,3,5 are shown in Figures \ref{gds_100_psf}, \ref{stack_2_psf} and \ref{stack_3_psf}. 

\section{Simulation}
\subsection{Simulating the Performance of the CZT MeV CubeSat}
We used the MEGAlib (Medium Energy Gamma-ray Astronomy Library)~\citep{MEGAlib} toolkit to perform simulations of all the configurations described above. Simulations of the effective area, angular resolution and energy resolution were carried out using monochromatic point sources with energies $0.1 \leq E \leq 4$ MeV, subject to cosmic- and albedo photon background radiation with the following energy spectra:
\begin{equation}
\label{Bkgspectrum}
        \Big(\frac{dN}{dE}\Big)_\text{albedo} = c_1 E^{-\Gamma_1},\
        \Big(\frac{dN}{dE}\Big)_\text{cosmic} = c_2 E^{-\Gamma_2},
\end{equation}
where $c_1 = 0.329\ \si{ph\ cm^{-2}\ s^{-1}\ keV^{-1}\ sr^{-1}}$, $\Gamma_1 = 1.34$ (for 868 keV - 20 MeV); $c_1 = 858\ \si{ph\ cm^{-2}\ s^{-1}\ keV^{-1}\ sr^{-1}}$, $\Gamma_1 = 2.12$ (for 20 MeV - 97.4 GeV); and 
$c_2 = 11.3\ \si{ph\ cm^{-2}\ s^{-1}\ keV^{-1}\ sr^{-1}}$, $\Gamma_2 = 2.15$ (for 868 keV - 107 GeV). These values were obtained by fitting background spectrum data from the MEGAlib Github  repository\footnotemark[4] \footnotetext[4]{\url{https://github.com/zoglauer/megalib/tree/main/resource/examples/cosima/source}}. The continuum sensitivity was simulated for sources whose energy spectra follow a power law $E^{-\Gamma}$ with spectral index $\Gamma = 1$ where the continuum source energy was taken up to 10 MeV. All simulations were performed for an effective observation time $T_{\text{eff}} = 10^{6}$ s, using a certain angular acceptance window $\alpha_\text{acc}$ within the energy peak (for point sources) or -range (for continuum sources) of the source's energy spectrum. 

\subsection{Energy Resolution}\label{sec:enres}
Energy resolution $R$ is a measure of how well a radiation detector can measure the energy of the incident source photons. In this paper, energy resolution was calculated as follows:
\begin{equation}
\label{ereseq}
    R = \frac{2\sigma}{E_0} \times 100\%,
\end{equation}
where $E_0$ is the energy of the source. The parameter $\sigma$ was found by fitting a normal distribution to the energy spectrum peak of the source and calculating the standard deviation of the fit. The energy spectrum was plotted within an acceptance window $w_\text{acc} = E_0 \pm 0.2E_0$ using 500 histogram bins. 

\subsection{Angular Resolution}\label{sec:arm}
The angular resolution of the CubeSat is defined by the \textit{angular resolution measure} (ARM), which, for Compton telescopes, is defined as the smallest angular distance between the incident direction of the $\gamma$-ray emitted by the source and each event \say{cone} traced out by the Compton scattered $\gamma$-ray. It gives a measure of how well the CubeSat payload can distinguish the true direction of the incident photon compared to the measured direction of the incident photon:
\begin{equation}
\label{armeq}
    \text{ARM} = \arccos{(\boldsymbol{\hat{e}}_g \cdot\boldsymbol{\hat{e}}_i)} - \varphi.
\end{equation}
Here, $\boldsymbol{\hat{e}}_g$ is the unit direction of the scattered $\gamma$-ray, $\boldsymbol{\hat{e}}_i$ the unit direction of the incident $\gamma$-ray, and $\varphi$ is the Compton scattering angle, defined in terms of the energies of the Compton scattered $\gamma$-ray ($E_g$) and electron ($E_e$) as:
\begin{equation}
\label{csa_eq}
    \cos{(\varphi}) = 1 - \frac{m_e c^2}{E_g} + \frac{m_e c^2}{E_g + E_e}.
\end{equation}
The distribution of the ARM is a representation of the point spread function (PSF) of the CubeSat, which is a measure of how well the CubeSat can resolve two neighboring point sources. The narrower the PSF, the better the ARM. In this paper, the quantity of interest for the analysis of the performance of the payload is $\sigma_\text{ARM}$, which denotes the Full Width at Half Maximum (FWHM) of the ARM PSF within $\alpha_\text{acc}$, where in this paper, $20^\circ \leq \alpha_\text{acc} \leq 50^\circ$. The angular acceptance window $\alpha_\text{acc}$ was chosen such that the full peak of the ARM PSF is contained without including too much of the \say{tails} of the PSF, therefore mostly taking into account Compton scattered gamma rays with correct kinematics and avoiding wrongly reconstructed events or incompletely absorbed electrons.

\subsection{Effective Area}
The effective area of a telescope is defined as its effective photon collection area (\say{aperture size}), and is often much smaller than its physical aperture size. It is calculated as follows:
\begin{equation}
\label{aeffeq}
    A_\text{eff} = \frac{N_\text{sel}}{N_\text{gen}} \cdot \pi r^2,
\end{equation}
where $N_\text{sel}$ is the number of events within $\alpha_\text{acc}$ after event selection and energy cuts in the \textit{Mimrec} MEGAlib library, $N_\text{gen}$ the total number of generated events in the \textit{Cosima} library, and $r$ the radius of the \say{surrounding sphere}\footnotemark[5]
\footnotetext[5]{See the Cosima manual for an explanation (\url{https://megalibtoolkit.com/documents/Zoglauer_Cosima.pdf}).}. The effective area can never become arbitrarily large because $N_\text{sel} \rightarrow 0$ as $r \rightarrow \infty$.

\subsection{MeV Continuum Sensitivity}
\label{sensitivity_section}
The telescope's ability to detect faint sources is an essential high-level performance measure that is quantified by the continuum sensitivity. To evaluate this sensitivity, the telescope's angular resolution, effective area, observation time, and background rate are all considered. In this study, the sensitivity has been computed for $\gamma$-ray sources at high Galactic latitude and high zenith angles concerning the Earth. 
Before presenting the sensitivity of the MeV CubeSat, a Compton telescope for $\gamma$-ray observation, a brief summary of the background sources for a Compton telescope in a low-Earth orbit is provided. The evaluation of the sensitivity requires a careful assessment of each background event source's contribution, as the Compton telescope is designed for a Low Earth Orbit (LEO; $\sim500$ km, equatorial orbital inclination). The main background contributions, as reported in \citep{2017ExA....44...25D, 2019ExA....47..273C}, are as follows:

\begin{description}
\item [Extra-galactic $\gamma$-ray background (EGB):] A photon background that is diffuse and isotropic is believed to stem from unresolved sources. This area of research is anticipated to become a significant subject of interest for Compton telescopes in the future~\citep{2019BAAS...51c.290A}.

\item [Earth’s $\gamma$-ray emission:] Earth's $\gamma$ emission is primarily generated by the interaction between primary cosmic rays and the Earth's atmosphere, which produces secondary particles and a bright $\gamma$-ray flux, also known as Earth’s albedo. Earth's albedo emission reaches its maximum at a polar angle of approximately $113^{\circ}$ from zenith when located at a low-Earth orbit altitude of 550 km. Even though no significant emission is expected for $<90^{\circ}$, the albedo background events of Earth still overlap with signal events in a Compton telescope as their origin can only be reconstructed to a great circle in the sky. The Earth's $\gamma$ emission contributes by far the most to the background, so the design of any MeV telescope requires work on strategies to minimize its impact.

\item [Charged-particle background:] The charged background refers to the events that are caused by the cosmic-rays that hit the detector. We note that a vast majority of these events ($\gtrsim 99.99\%$) can be effectively vetoed through the use of an anti-coincidence detector (ACD), and hence the contribution from the charged-particle background has been ignored in this study.

\item [Material activation:] The steady and unceasing influx of cosmic rays and $\gamma$-rays, which incessantly bombard the spacecraft, results in the activation of the materials of the satellite, thereby generating radioactive isotopes. The overall background rate during on-orbit operations is considerably influenced by the spacecraft's materials and payload, as well as its orbit. For small satellites like this MeV CubeSat payload, the instrumental background is expected to be lower than that observed for large-scale missions, and it scales roughly 1/3 times the mass ratio. For the assessment of the MeV CubeSat's sensitivity, material activation is disregarded in the first instance, but it should be taken into account in the future when all the payload and orbit details are concluded.
\end{description}

The continuum sensitivity can be estimated through a semi-analytical method. Background ﬂuxes are derived from previous missions' data: Earth's $\gamma$ emission from \citep{2017ExA....44...25D} and EGB from \citep{1999ARNPS..49..163R}, extended at lower energies\footnotemark[6] \footnotetext[6]{As the measured EGB ﬂux depends on resolved sources, and therefore for telescope sensitivity calculations COMPTEL’s data was used, which performed similarly to ours.}. For point-like $\gamma$-ray sources, the continuum sensitivity for the MeV CubeSat payload can be expressed by:

\begin{equation}
\label{ContSens}
    F_z = \frac{z^2 + z\sqrt{z^2 + 4N_B}}{2T_\text{eff}A_\text{eff}},
\end{equation}

where $z$ is the statistical significance in a unit of sigmas (here, $3\sigma$ source detection), $T_\text{eff}$ is the total observation time, $A_\text{eff}$ is the effective area, and $N_B$ is the number of background photons that lie within the angular resolution element defined by the telescope.
Then, the total number of background events is given by:
\begin{equation}
\label{NBackground}
    N_B \simeq N_\text{egb} \cdot (1 + R),
\end{equation}
where $R$ is the ratio between Earth’s $\gamma$ emission and EGB events, and $N_\text{egb}$ is the number of EGB events inside the angular resolution element. Hence, $N_B$ is calculated using:
\begin{equation}
\label{Nb}
    N_B = \Phi \cdot T_\text{eff} \cdot A_\text{eff} \cdot \Delta\Omega.
\end{equation}
Here, $\Phi$ is the background flux, which can be calculated using the energy spectra in Eq. \eqref{Bkgspectrum} as:
\begin{equation}
    \Phi = \int \sum_{i=1}^n \Big(\frac{dN}{dE}\Big)_i\, dE,
\end{equation}
and $\Delta\Omega$ is the angular resolution element:
\begin{equation}
    \Delta\Omega = [\cos(\overline{\varphi} - \sigma_\text{ARM}) - \cos(\overline{\varphi} + \sigma_\text{ARM})] \cdot 2\sigma_\text{SPD}.
\end{equation}

Here, $\varphi$ is calculated via Eq. \eqref{csa_eq} and $\sigma_\text{SPD}$ is the Half Width At Half-Maximum (HWHM) of the fitted scatter plane deviation (SPD) distribution for the scattered electron and can be calculated in terms of the unit direction of the scattered $\gamma$-ray, the initial $\gamma$-ray, and the electron as follows:

\begin{equation}
\label{SPDcalc}
    \text{SPD} = \arccos{[(\boldsymbol{\hat{e}}_g \times \boldsymbol{\hat{e}}_i) \cdot (\boldsymbol{\hat{e}}_g \times \boldsymbol{\hat{e}}_e)]}.
\end{equation}

The CZT crystal modules studied here do not possess the capability of tracking electrons and thus, there exists no measure for $\boldsymbol{\hat{e}}_e$. Due to this reason, in the analysis, $\sigma_\text{SPD}$ was established as 180$^{\circ}$. The simulation also considers an assumption that the detector is directed towards the zenith, which is generally away from the Earth and the source is positioned at the zenith angle $\theta = 0$, which is the most optimal situation for the sensitivity value. We note that the ﬂux of $\gamma$-rays from the Earth is essentially zero at angles $\theta<90^{\circ}$. Nonetheless, a minute proportion of events presents a sizable SPD, which means that the Compton arc is partially present in the sky map element that is being observed. Since the $\gamma$ emission ﬂux of the Earth is significantly more at LEO, it is still the most significant background source for the payload design. The MeV CubeSat payload has the potential to cover the energy range of roughly 0.1 MeV to 10 MeV with a similar sensitivity level compared to the previous generation instruments of large-scale MeV missions. The larger field of view and much lower cost are fundamental for the observation of transients and $\gamma$-ray bursts, where sky coverage plays a pivotal role.

\section{Results}\label{results}
\begin{figure*}[h]
\begin{minipage}{\textwidth}
\centering
\subfigure[Effective area]{\label{aeff_gds}
\includegraphics[width=0.48\textwidth,height=0.22\textheight]{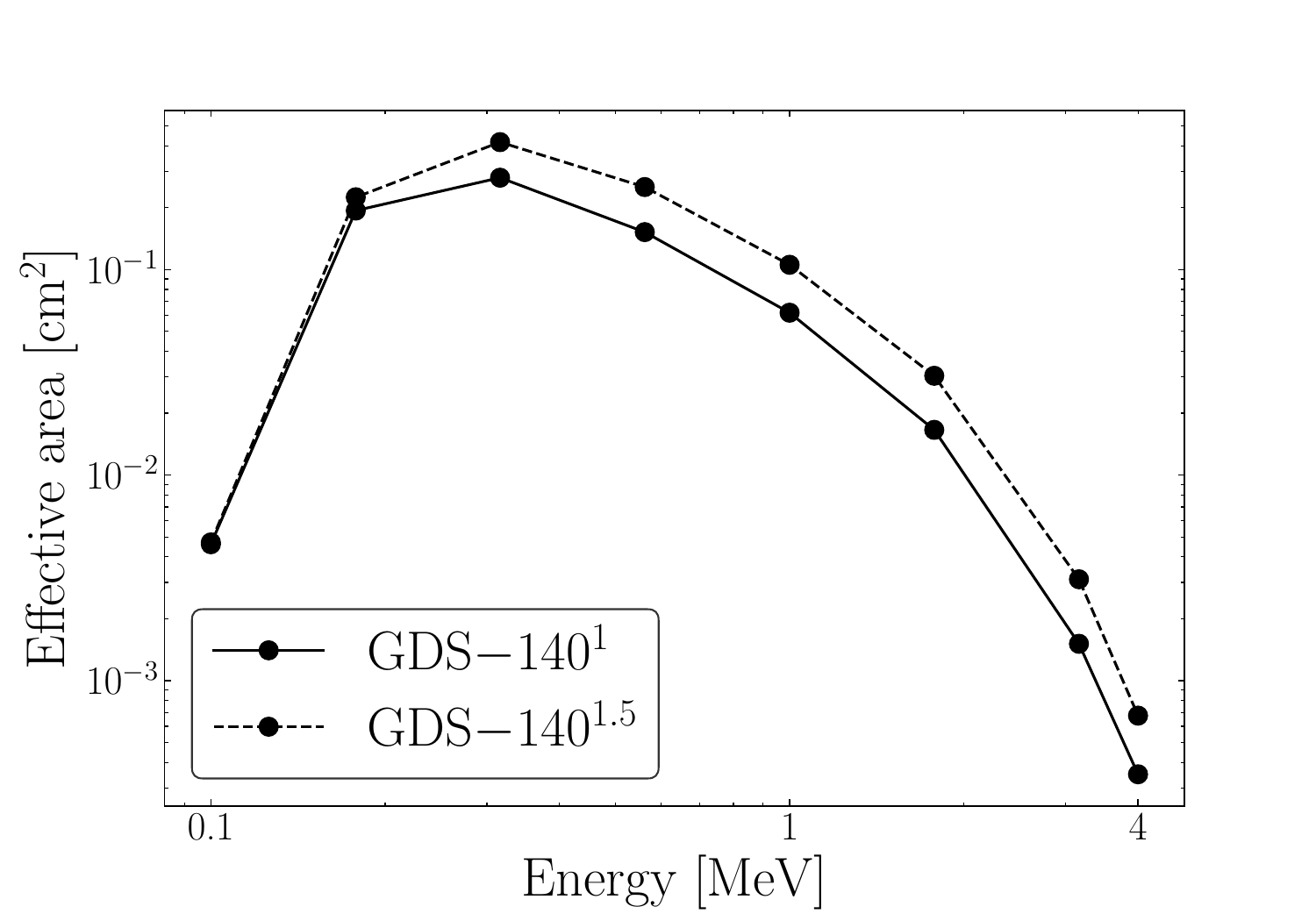}} 
\subfigure[Angular resolution]{\label{arm_gds}
\includegraphics[width=0.48\textwidth,height=0.22\textheight]{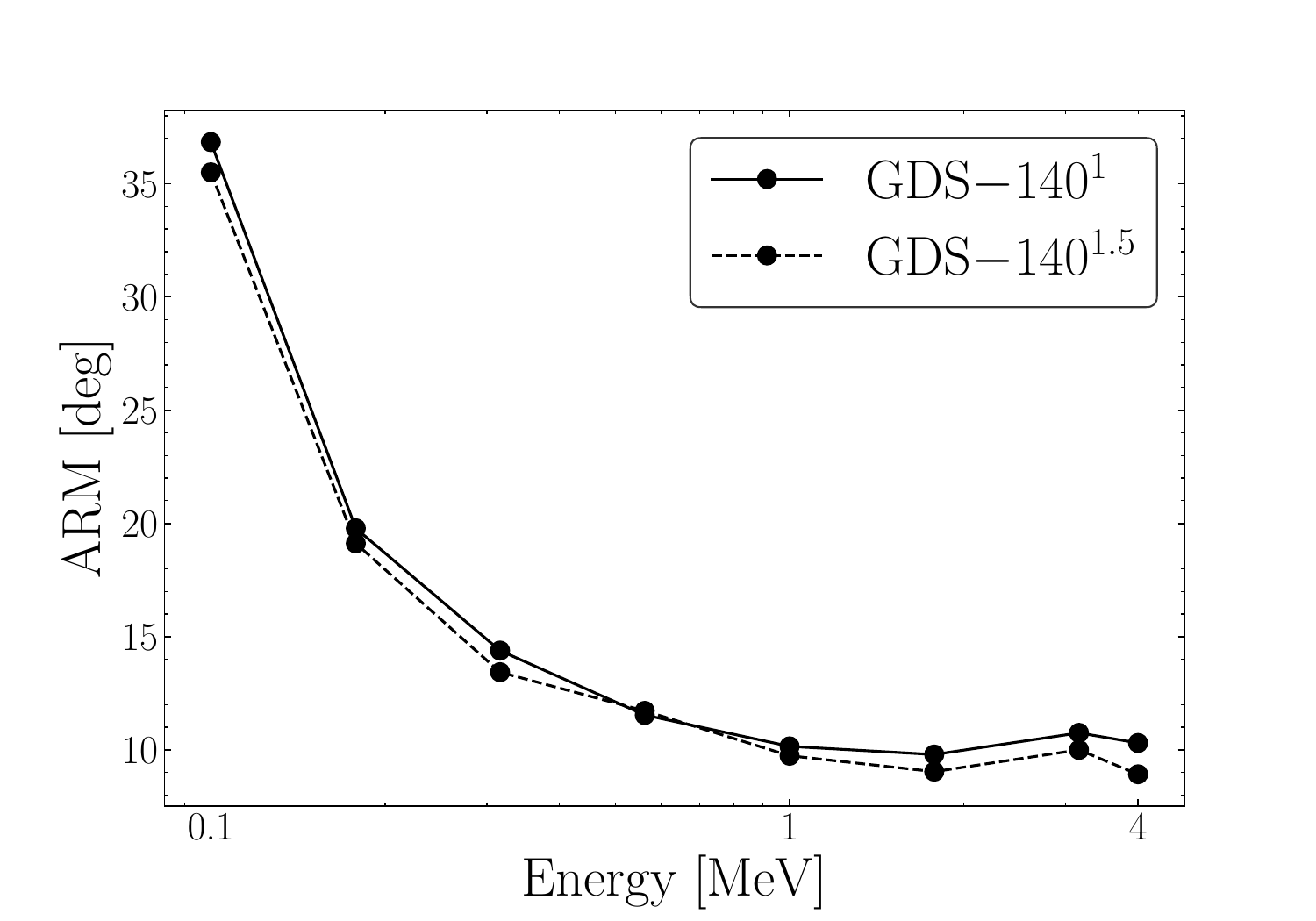}}
\subfigure[Energy resolution]{\label{eres_gds}
\includegraphics[width=0.48\textwidth,height=0.22\textheight]{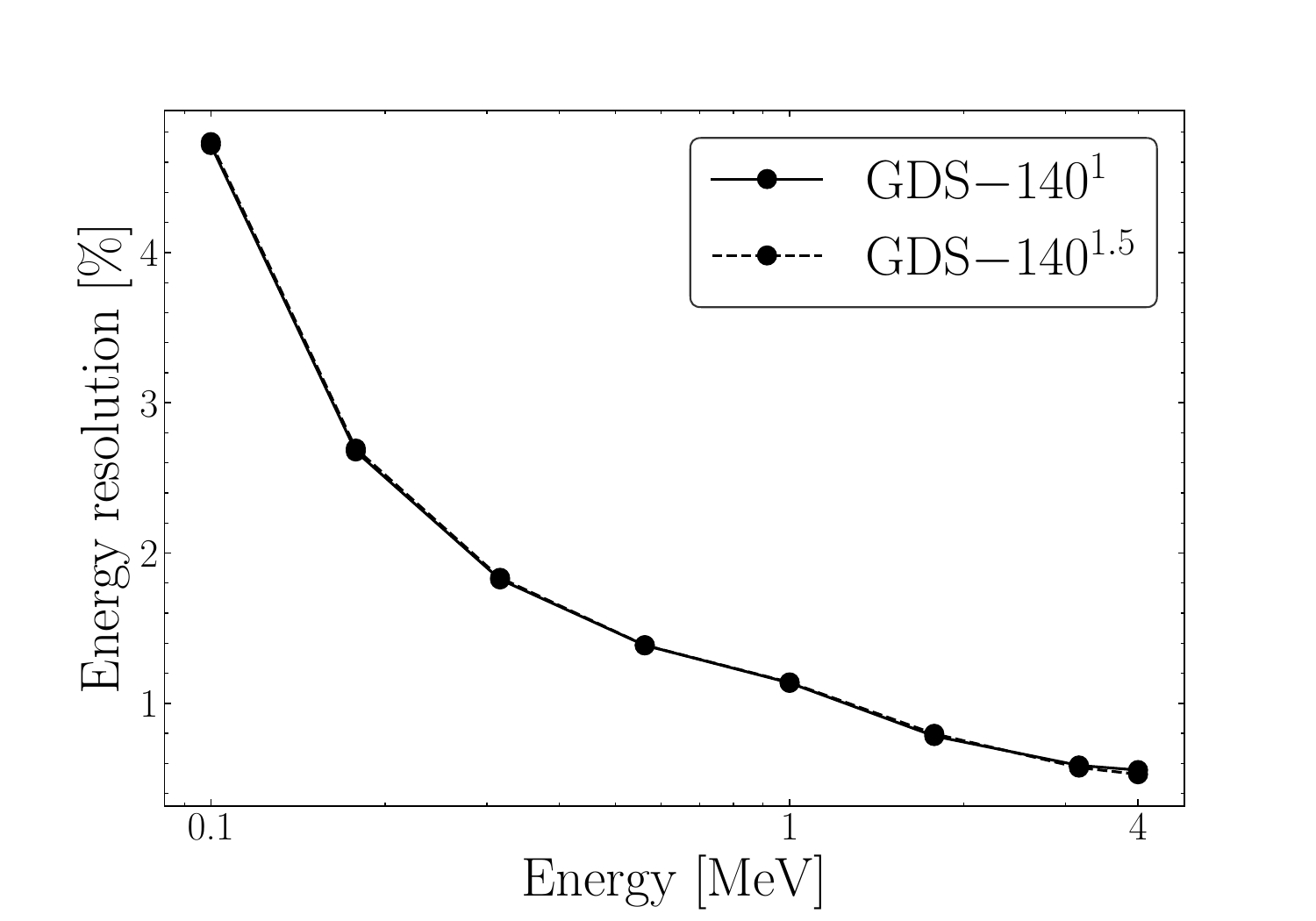}}
\subfigure[Sensitivity]{\label{sensi_gds}
\includegraphics[width=0.48\textwidth,height=0.22\textheight]{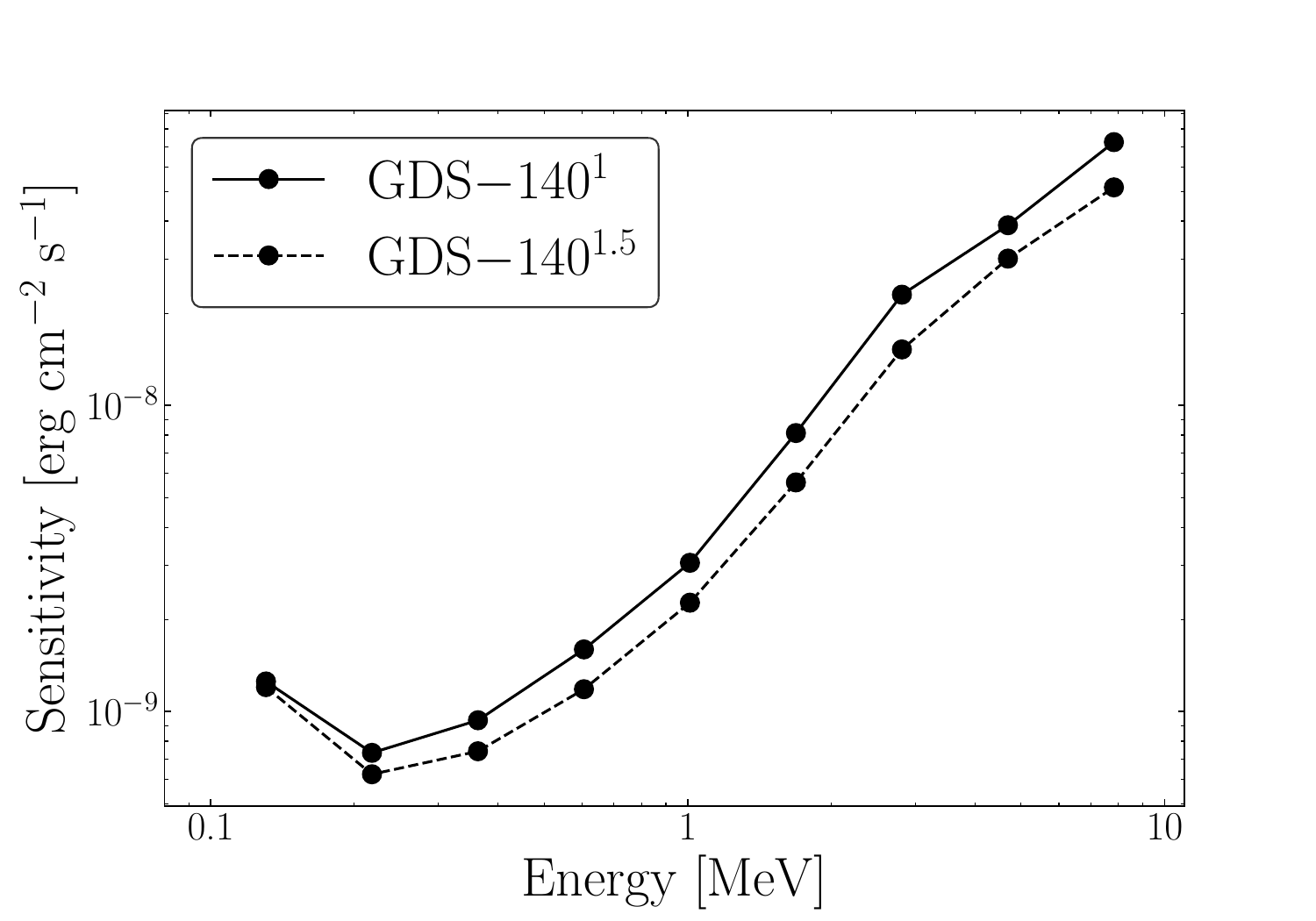}}
\caption{Comparison of the simulated effective area, angular resolution, energy resolution, and continuum sensitivity with crystal widths $d=1$ cm and $d=1.5$ cm for the standard \amdl\ designs. The solid line represents \mgds, and the dashed line represents \mgdsthick.} 
\label{gds_comparison}
\end{minipage}
\end{figure*}

Figure \ref{gds_comparison} shows a performance comparison between \mgds\ and \mgdsthick\ for the effective area, angular/energy resolution and sensitivity. 
It is observed that \mgdsthick\ achieves better effective area and angular resolution from 0.1 to 4 MeV compared to \mgds, which is due to a larger amount of reconstructed events arising from the larger Compton cross section for \mgdsthick. By the same argument, \mgdsthick\ achieves better ARM compared to \mgds. At energies below 0.2 MeV, the effective area difference between \mgds\ and \mgdsthick\ is very small due to the photoelectric effect being the main interaction process there, which also explains the lower effective area at these energies.

The decreasing trend in the effective area beyond 0.3 MeV arises because more Compton scattered electrons with tracks (undetectable by the CZT crystal) are produced as the energy of the incident $\gamma$ rays increases.

Figure \ref{eres_gds} illustrates how the energy resolution does not depend on the crystal thickness, but is only dependent on the type of detector material being used and the gain voltage between the anode and cathode connected to the detector. Experiments done by IDEAS achieve a \amdl\ energy resolution of 0.76\% FWHM for a Cs-137 source at 0.662 MeV, although our simulated energy resolution predicts a FWHM of around 1.2\% at this same energy. This discrepancy most likely arises because MEGAlib does not have the capability to simulate electronics and the corresponding gain voltages. It should therefore be noted that the reported results in this paper only give rough estimates of the true performance of a \amdl, as the performance may either increase/decrease depending on the gain voltage. 

In Figure \ref{sensi_gds}, the sensitivity increases significantly for \mgdsthick\ compared to \mgds, except at energies less than 0.2 MeV due to the photoelectric effect.

\begin{figure*}[h]
\begin{minipage}{\textwidth}
\centering
\subfigure[Effective area]{\label{aeff_mod123}
\includegraphics[width=0.48\textwidth, height=0.22\textheight]{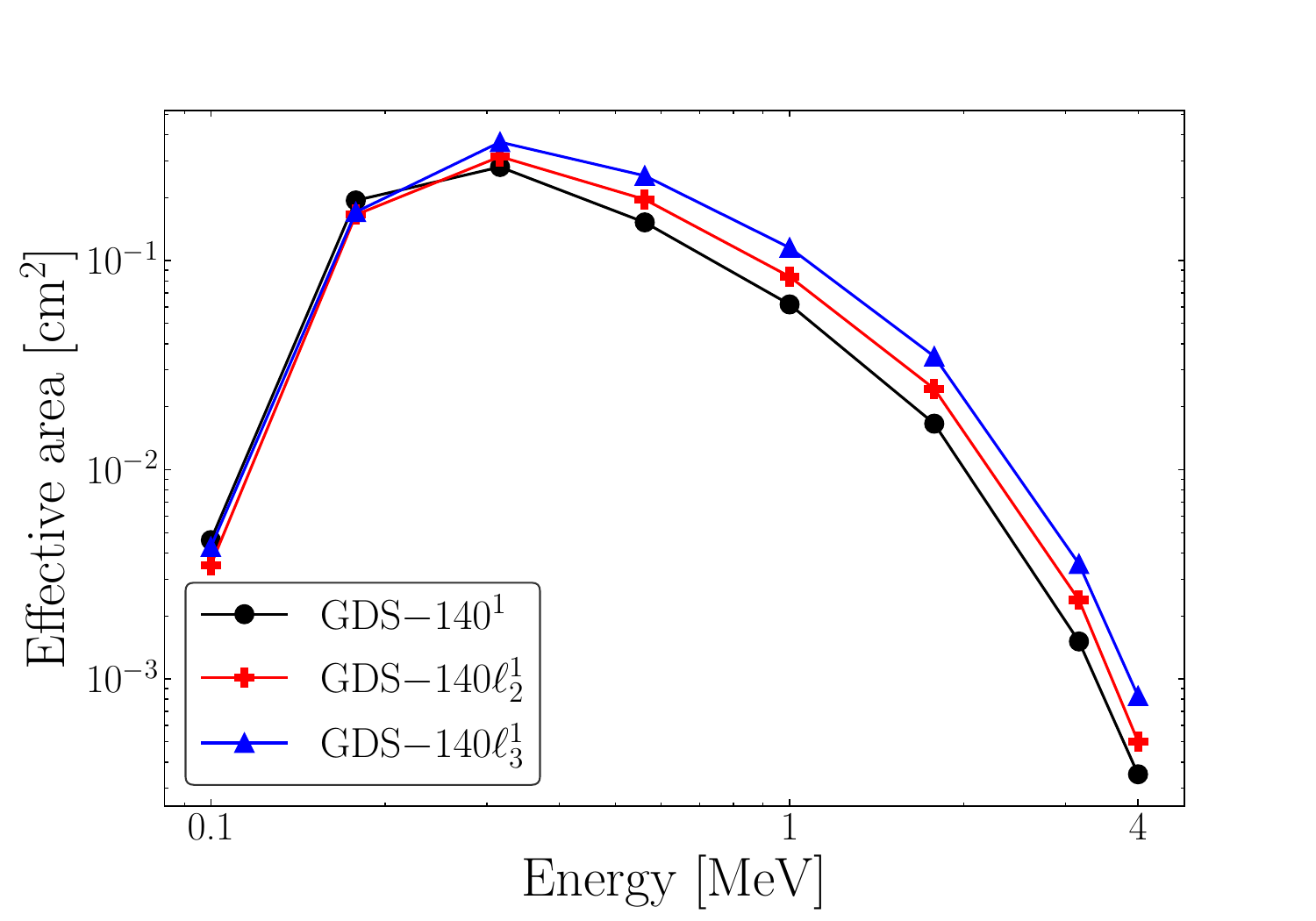}}
\subfigure[Angular resolution]{\label{arm_mod123}
\includegraphics[width=0.48\textwidth, height=0.22\textheight]{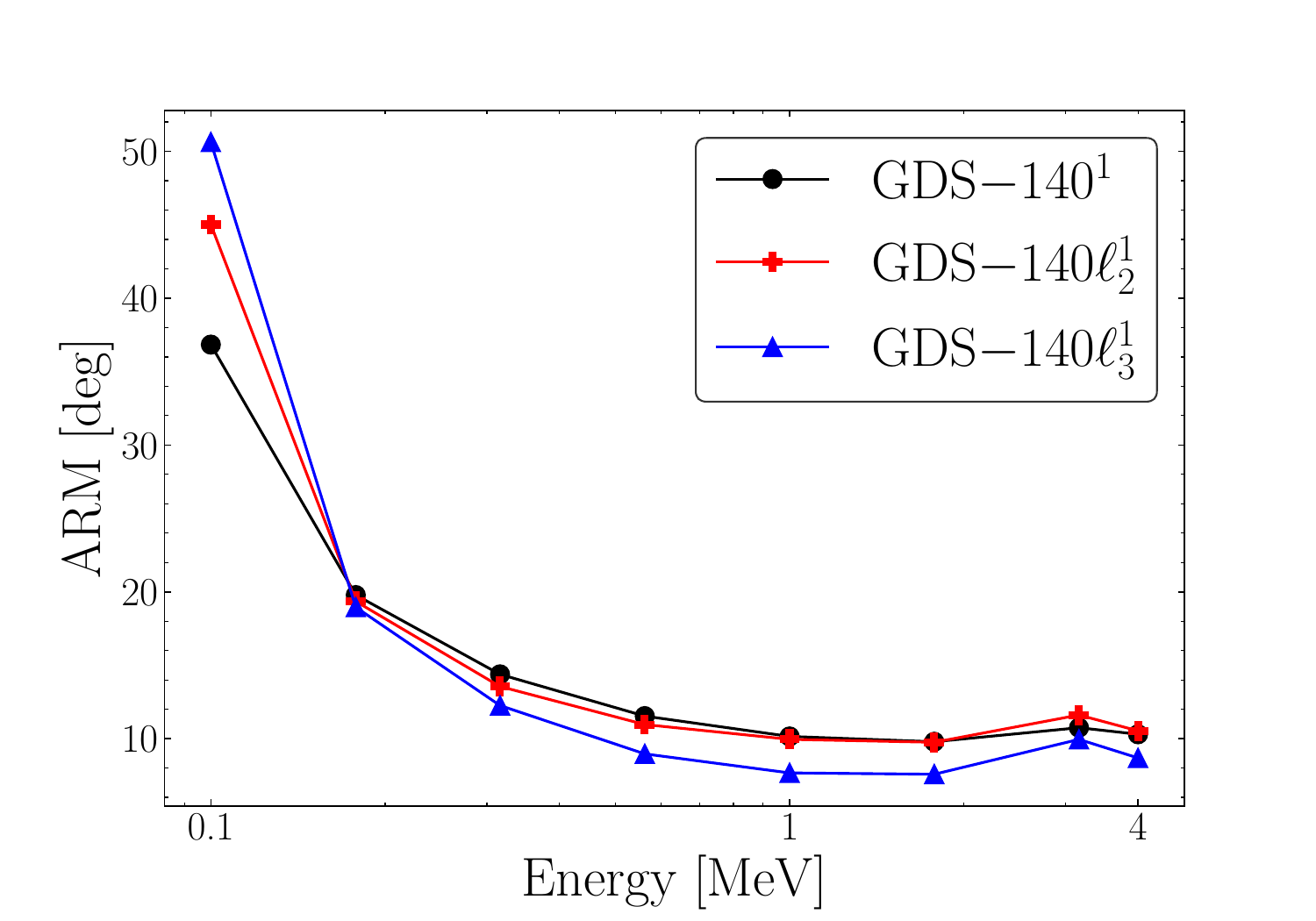}}
\subfigure[Energy resolution]{\label{eres_mod123}
\includegraphics[width=0.48\textwidth, height=0.22\textheight]{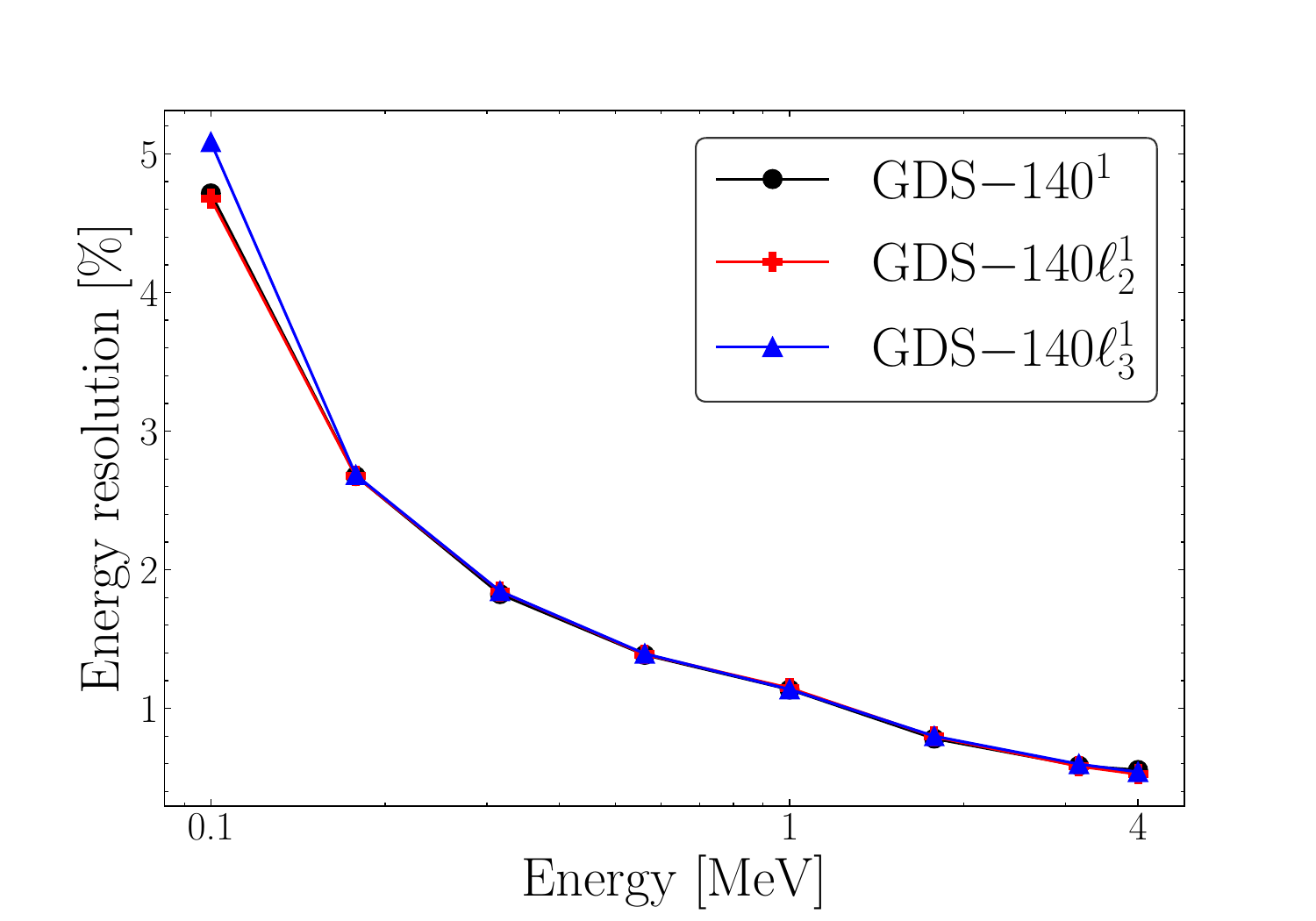}}
\subfigure[Sensitivity]{\label{sensi_mod123}
\includegraphics[width=0.48\textwidth, height=0.22\textheight]{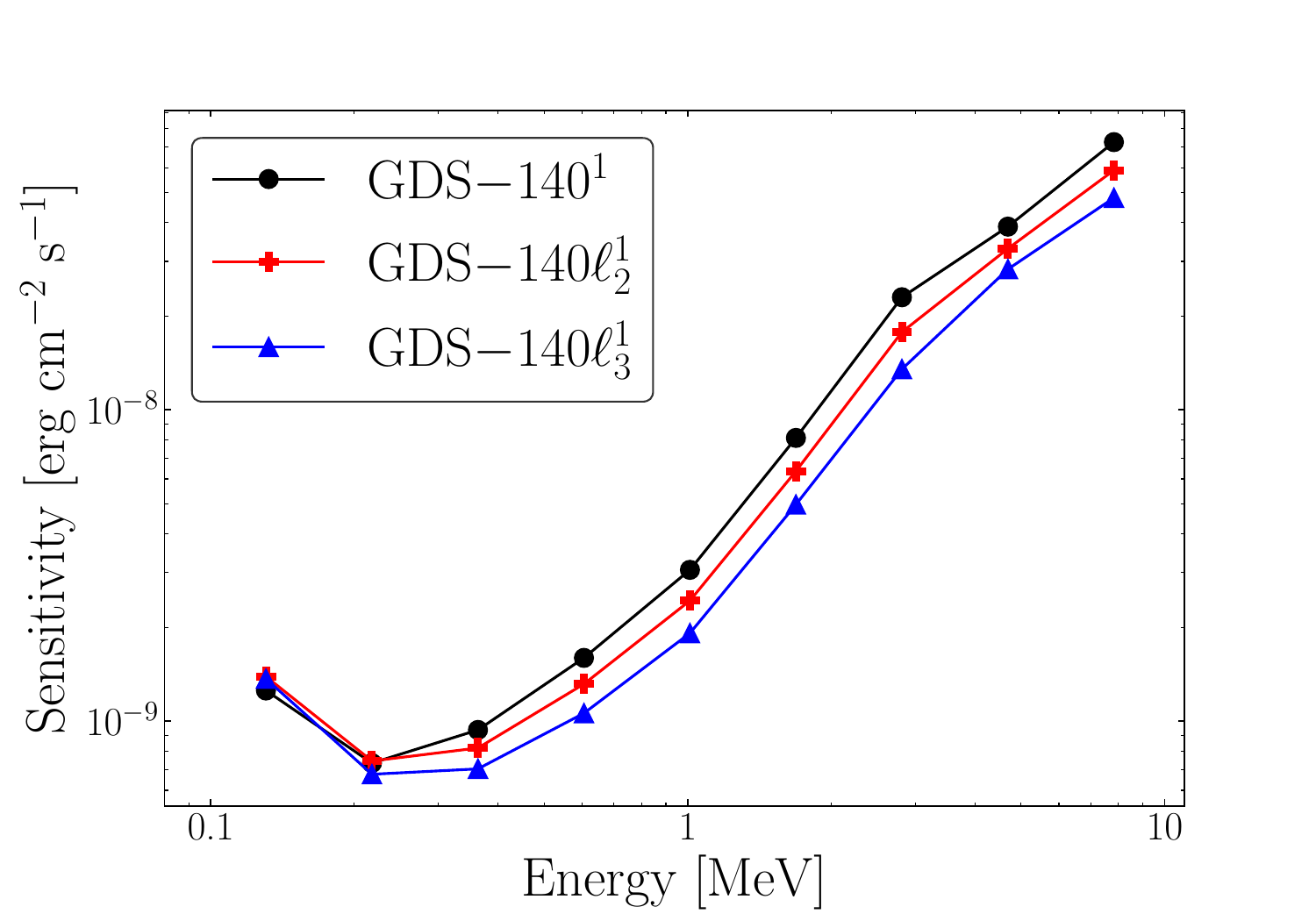}}
\caption{Simulated effective area, angular resolution, energy resolution and continuum sensitivity comparison for \amdl\ payload designs incorporating different $d=1$ cm crystal layers. The black line and dots represent \mgds\, the red line and crosses represent \mstack\ and the blue line and triangles represent \mstackthree.} 
\label{gds_123}
\end{minipage}
\end{figure*}

Whereas in Figure~\ref{gds_comparison}, the performance difference for different crystal thicknesses is shown, the performance difference for different crystal layers is shown in Figure~\ref{gds_123} for \mgds, \mstack\ and \mstackthree, keeping the crystal thickness constant. 
Next to the same trend (with increasing energy) as in Figure~\ref{gds_comparison}, it can furthermore be noticed that when the amount of crystals is increased (e.g., by stacking up multiple layers), the effective area, angular resolution and sensitivity increases. One layer achieves the worst performance, whereas three layers achieve the best performance beyond the photoelectric-dominant energy region in the analysis. 

\begin{figure*}[h]
\begin{minipage}{\textwidth}
\centering
\subfigure[Effective area]{\label{aeffnorm_mod123}
\includegraphics[width=0.48\textwidth, height=0.22\textheight]{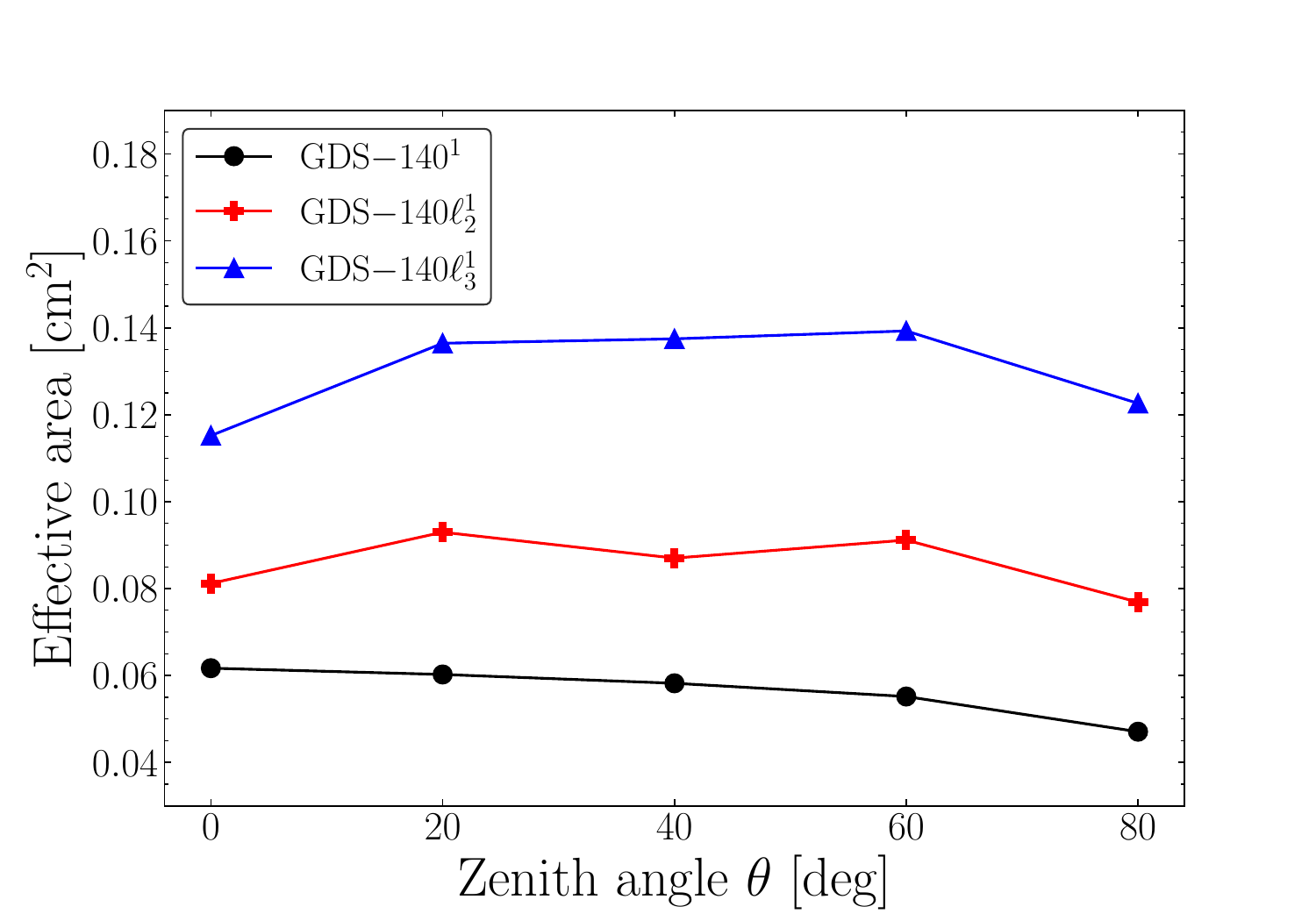}}
\subfigure[Angular resolution]{\label{armzenith_mod123}
\includegraphics[width=0.48\textwidth, height=0.22\textheight]{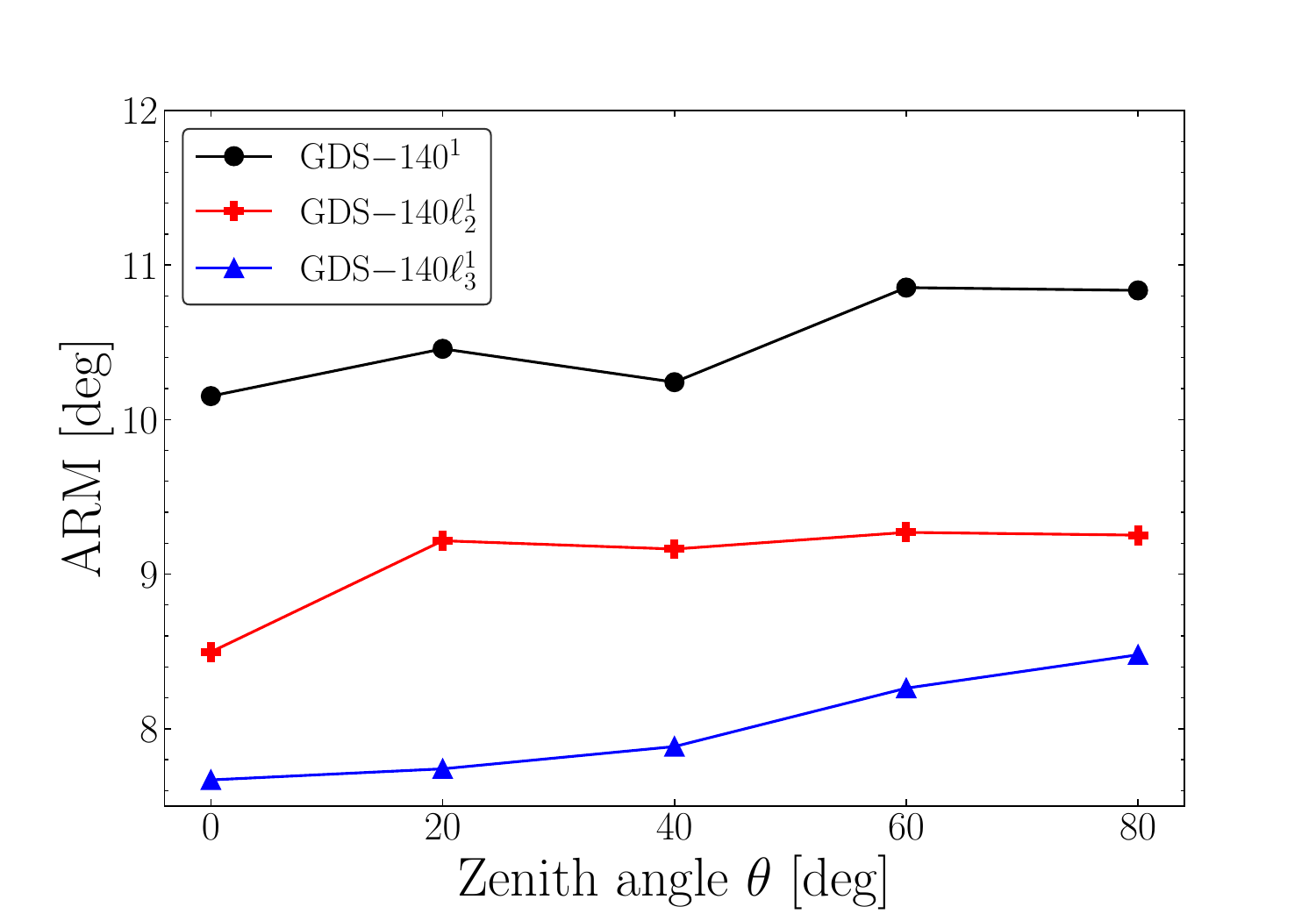}}
\subfigure[Energy resolution]{\label{ereszenith_mod123}
\includegraphics[width=0.48\textwidth, height=0.22\textheight]{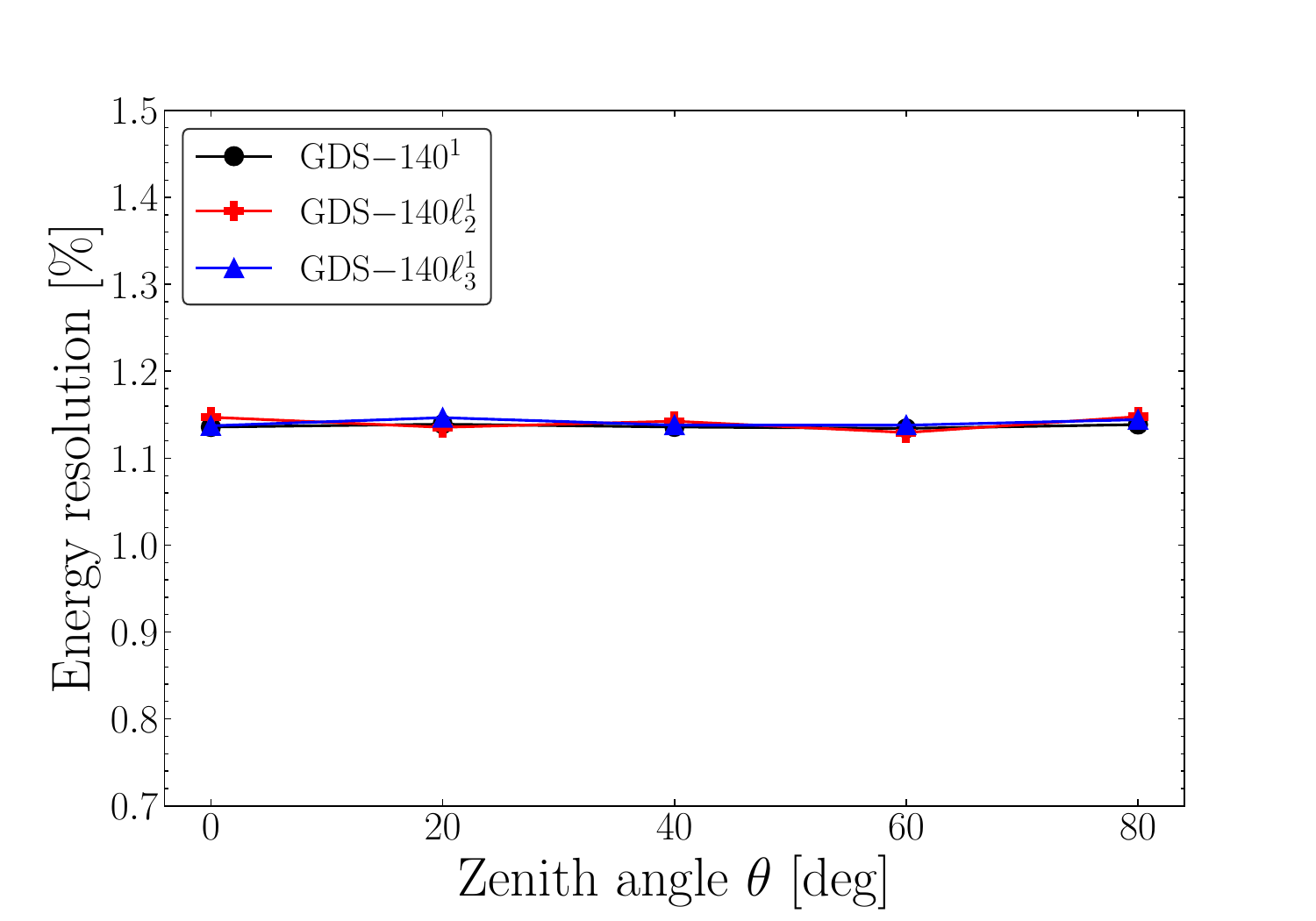}}
\subfigure[Sensitivity]{\label{fzzenith_mod123}
\includegraphics[width=0.48\textwidth, height=0.22\textheight]{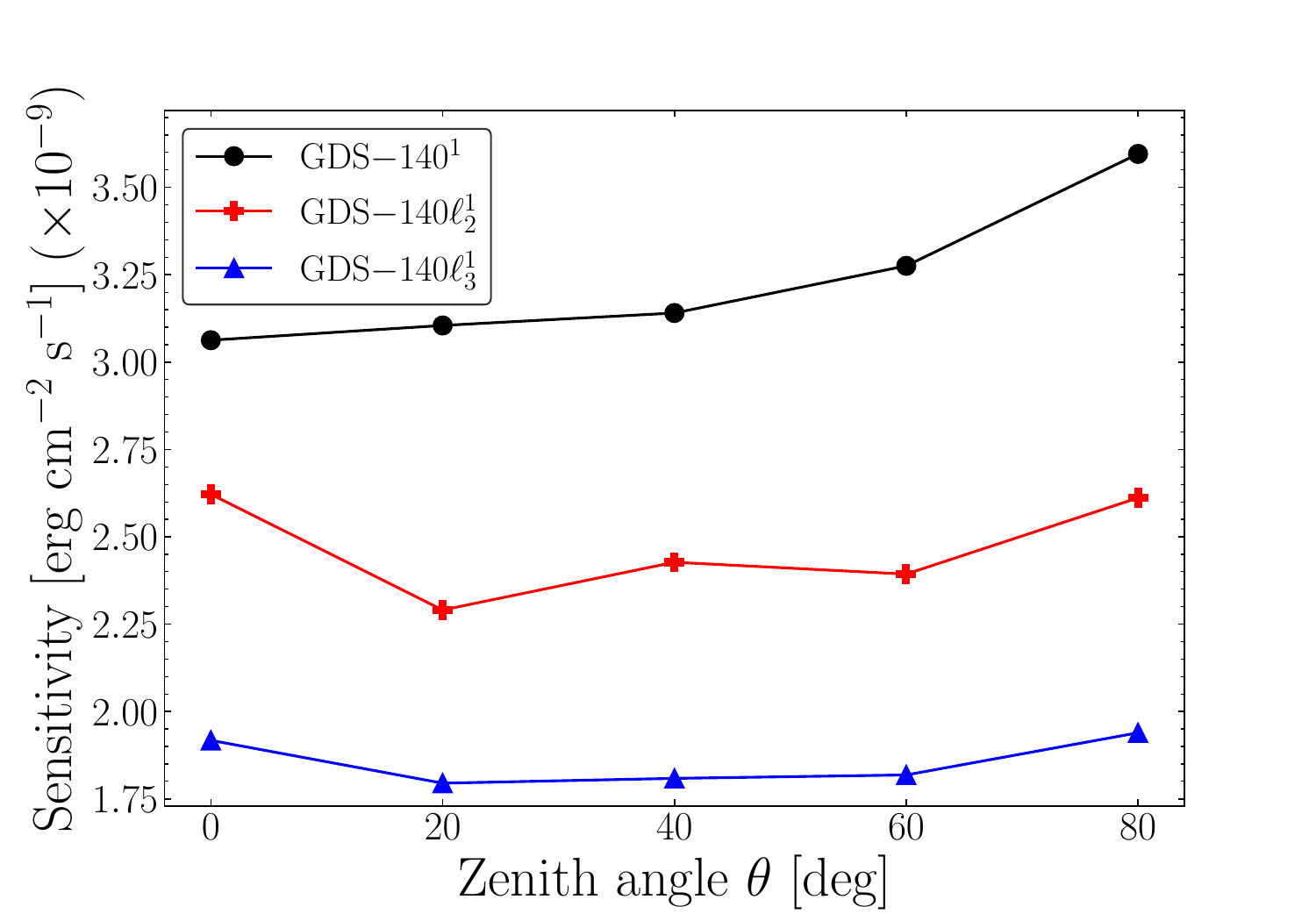}}
\caption{Simulated effective area, angular resolution, energy resolution, and continuum sensitivity comparison for \mgds, \mstack\ and \mstackthree\ with varying zenith angle $\theta$ at 1 MeV. In subfigure \ref{ereszenith_mod123}, the energy resolution axis was scaled such that the small variations ($\mathcal{O}(10^{-2})$) in the energy resolution do not look too large compared to the calculated energy resolution values. The angular acceptance window was kept the same for all varied zenith angles $\theta$.}
\label{gdszenith_123}
\end{minipage}
\end{figure*}

Next to varying the photon energy, a comparison of the effective area, angular/energy resolution, and sensitivity was also made for varying the source zenith position angle $\theta$, where $\theta$ was varied from $0^{\circ}$ to $80^{\circ}$ in steps of $20^{\circ}$ at 1 MeV. The results are shown in Figure~\ref{gdszenith_123}. 
In Figure~\ref{aeffnorm_mod123}, the effective area shows variations without a particular trend, although variations are most significant for \mstackthree.
For the angular resolution, shown in ~\ref{armzenith_mod123}, a slightly increasing trend can be noticed as $\theta$ increases. 
In Figure~\ref{ereszenith_mod123}, the energy resolution shows no clear trend because the energy resolution is dependent on the material and not on the energy or position of the source. Slight variations in the energy resolution may occur because the energy window containing the energy spectrum peak is set manually, therefore small fitting parameter value differences for each angle may occur. Regarding the variations in sensitivity shown in Figure \ref{fzzenith_mod123}, no clear trend can be observed as well.  

\begin{table}[ht]
\caption{Simulated detector performance comparison at 1 MeV with previous detector technology designs. Sensitivity is given in units of $10^{-9}$\fzu. L22a and MASS-Cube parameters are adapted from \cite{2022JCAP...08..013L} and \cite{mass2023} respectively.} 
\label{geo_table}
\begin{tabular*}{\textwidth}{@{\extracolsep\fill}ccccccc} %
\toprule
Payload & Volume & CZT Crystals & Effective & ARM & Energy& Sensitivity \\
designs & (U) & (\#) & Area ($\si{cm^2}$) & ($^{\circ}$) & Resolution(\%)& ($T_\text{eff}$) \\
\midrule 
LSR-1 & 9 & 128 (d=1.5~cm) & 1.61 & 5.58 & 1.16 & 1.18~($10^5$\ s) \\ 
L22a & 4 & 128 (d=1.5~cm) & 11.16 & 1.68 & 2.36 & 0.86~($10^5$\ s) \\
\midrule
LSR-2 & 4 & 32 (d=1.5~cm) & 0.40 & 6.89 & 1.17 & 2.71~($10^5$\ s) \\ 
L22a & 1 & 32 (d=1.5~cm) & 2.05 & $\dotsm$ & $\dotsm$ & 2.39~($10^5$\ s) \\ 
\midrule
\mgdsthick & 1 & 4 (d=1.5~cm) & 0.11 & 9.74 & 1.2 \footnotemark[7] & 2.27~($10^6$\ s) \\
MASS-Cube & 1 & 4 (d=1.5~cm) & 0.15 & 5.9 & 0.6 \footnotemark[7] & 5.05~($10^6$\ s) \\
\botrule
\end{tabular*}
\footnotetext[7]{Energy resolutions of MASS-Cube and \mgdsthick are calculated for 0.662 MeV.}
\end{table}

\section{Discussion}
Table~\ref{geo_table} shows the comparison between the simulated performance of the newly developed CZT module by IDEAS, MASS-Cube \citep{mass2023} and the old models \citep{2022JCAP...08..013L, 2022JInst..17P8004L}. In \cite[see Figure~2]{2022JCAP...08..013L}, a simulated payload design with 128 CZT crystals (i.e. $2 \times 2 \times 1.5\ \si{cm^3}$) within a 4U payload design \citep[see Table~1]{2022JCAP...08..013L} was proposed. Here, a similar payload design (LSR-1), with 32 modules containing 128 CZT crystals of the same size, is simulated. The material properties (i.e., energy resolution and depth resolution of CZT crystal) for simulation are obtained from Redlen Technologies. A similar comparison with another payload design, with 32 crystals from \cite{2022JCAP...08..013L}, is also compared with a similar payload design (LSR-2) using 8 modules here. During the comparison of simulated parameters, the old simulated design outperforms the new design in the effective area and ARM. The energy resolution is better in the new design and the sensitivity values are comparable with the old design. But the old design motherboard (VATA-450.3) from IDEAS is obsolete now and has been replaced with the new motherboard (i.e., GDS-100).

This comparison study is an effort to compare the simulated performance of the new detector design with the old design, although direct comparison is not possible due to technical limitations. The CZT crystal readout for the new design is obtained using ($11 \times 11$) pixels over the old readout design of ($8 \times 8$) pixels \citep[see Figure~1]{2022JInst..17P8004L}. In the old payload design simulations \citep[see Figure~2]{2022JCAP...08..013L}, the CZT crystals are placed in a tight packing, whereas in the new design, CZT crystals are placed with an offset of 12 mm (see Figure~\ref{payload_design} here). Due to the introduction of offset between the crystals, the payload volume in the latest module design is more than in the previous version.

While \cite{2022JCAP...08..013L} explores simulations of payload designs using numerous CZT crystals, it only presents lab tests for a single CZT crystal detector, as detailed in \cite{2022JInst..17P8004L}. Moreover, there is no existing report on detector designs employing multiple (VATA450.3) motherboards. The designs we simulate here take into account both the motherboard and the high-voltage power source volume, offering a more practical approach in terms of application, and providing a closer approximation to the actual volume needed for the real payload. This makes our payload design more aligned with the realities of payload manufacturing.

In \cite{mass2023}, a simulated payload design featuring CZT crystals, named MASS and MASS-cube, is presented. In Table~\ref{geo_table}, the simulated performance of MASS-Cube and \mgdsthick\ are compared. MASS-Cube shows marginally superior parameters compared to GDS-140 (except for sensitivity), although these differences might stem from varying simulation algorithms and assumptions.

Future laboratory experiments will involve using a \amdl\ with different gain voltages to validate the preliminary simulation results presented in this paper. The findings from these experiments will form the basis of a forthcoming paper.

\begin{figure*}[h]
\begin{minipage}{\textwidth}
\centering
\subfigure[Effective area]{\label{aeff_gds1MeV}
\includegraphics[width=0.48\textwidth, height=0.22\textheight]{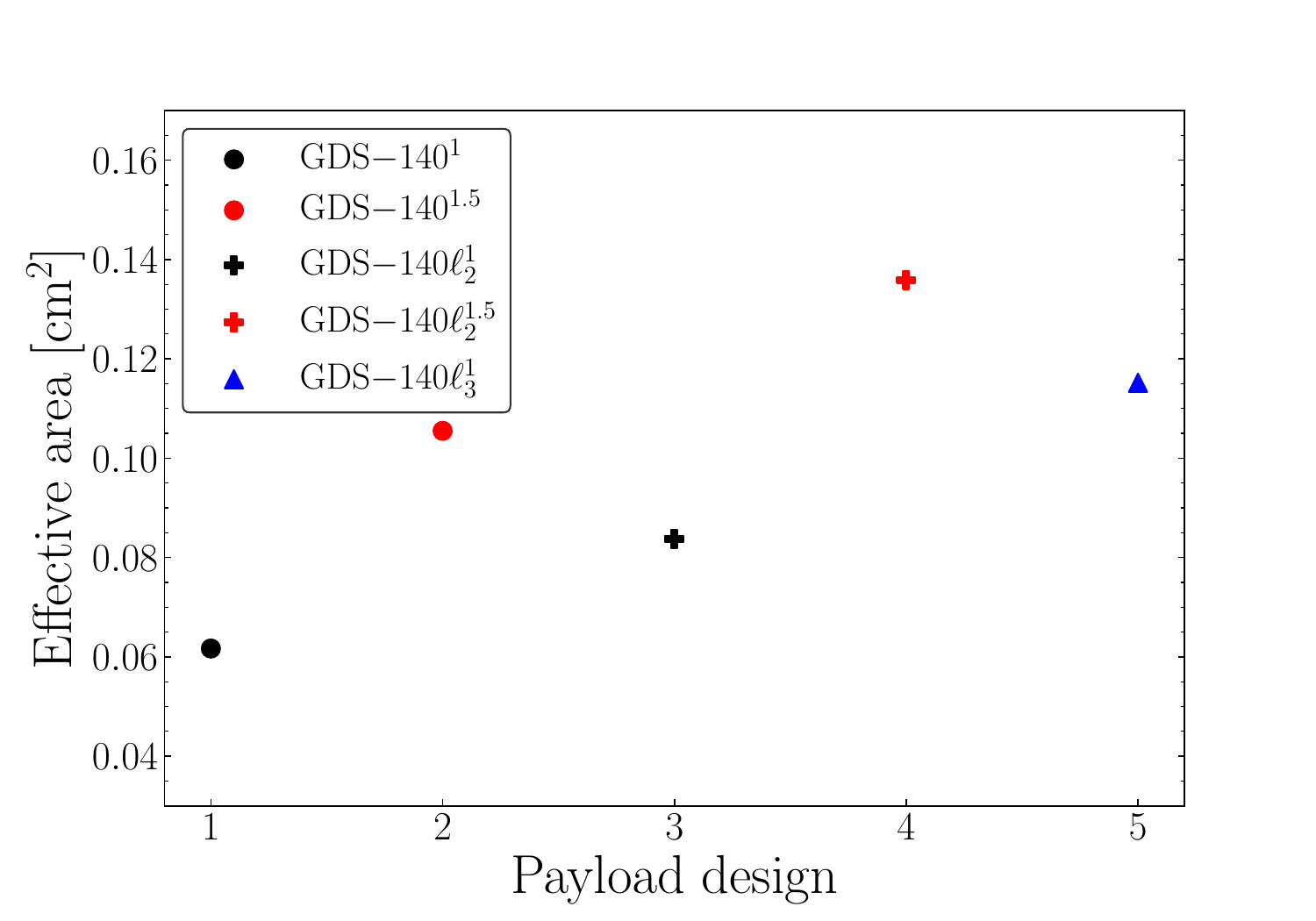}}
\subfigure[Angular resolution]{\label{arm_gds1MeV}
\includegraphics[width=0.48\textwidth, height=0.22\textheight]{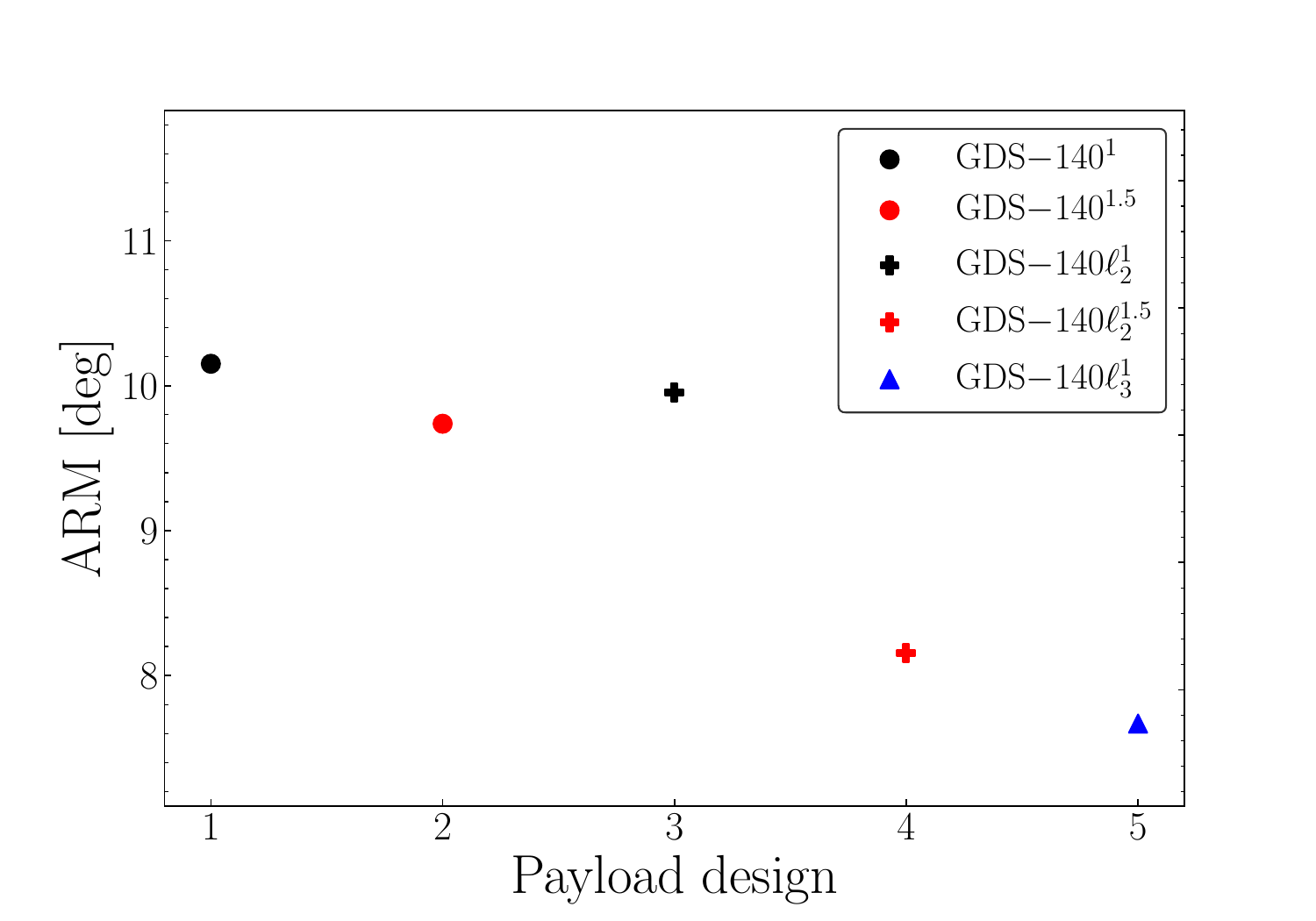}}
\subfigure[Energy resolution]{\label{eres_gds1MeV}
\includegraphics[width=0.48\textwidth, height=0.22\textheight]{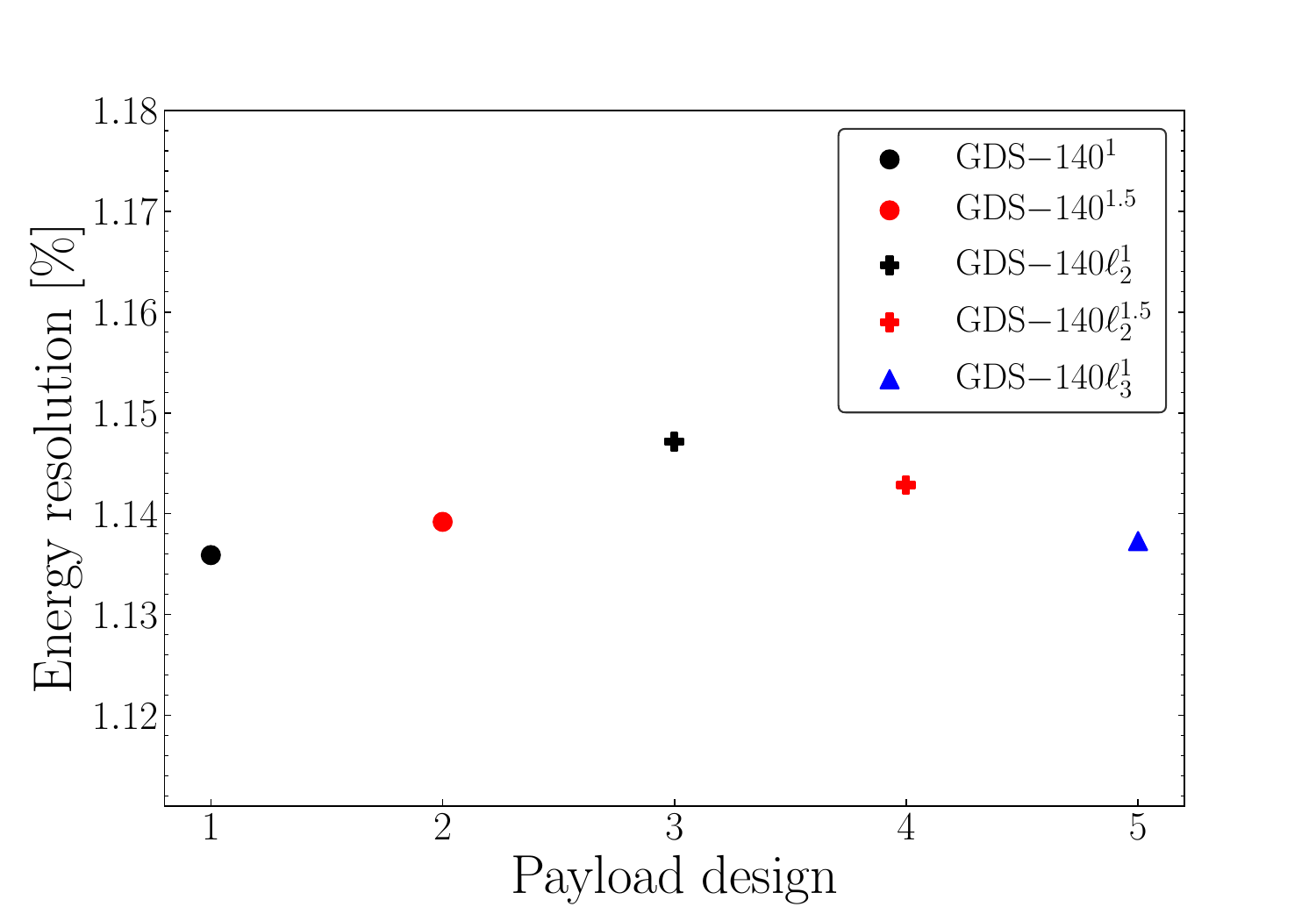}}
\subfigure[Sensitivity]{\label{sensi_gds1MeV}
\includegraphics[width=0.48\textwidth, height=0.22\textheight]{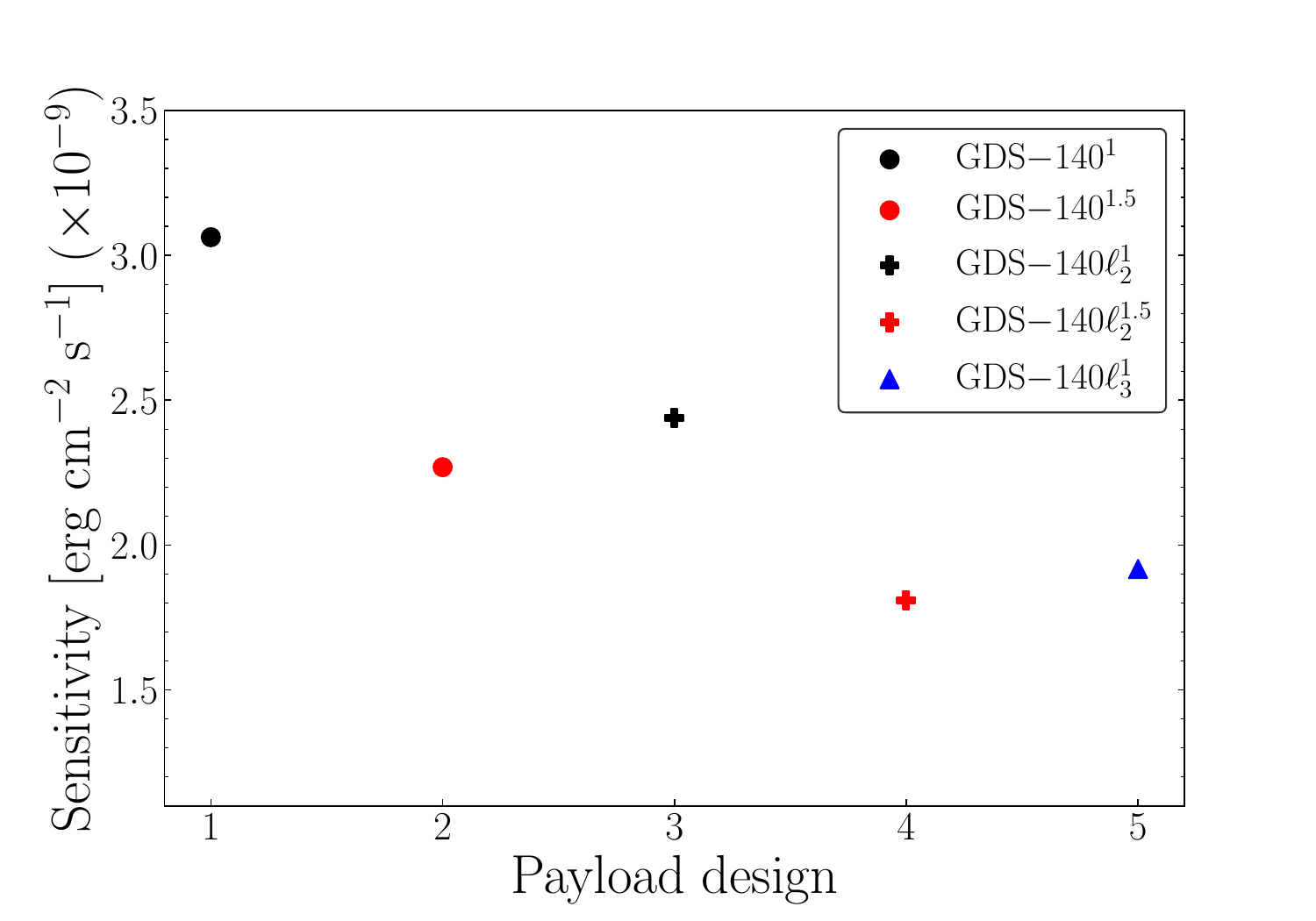}}
\caption{Simulated effective area, angular and energy resolution, and continuum sensitivity comparison for all \amdl\ designs at 1 MeV. Black and blue data points represent a crystal depth $d = 1$ cm, and red data points represent a crystal depth $d=1.5$ cm. Different data point symbols were used, where dots represents 1 crystal layer, crosses represent 2 crystal layers, and triangles represent 3 crystal layers.}
\label{gds_comparison1MeV}
\end{minipage}
\end{figure*}

In previous sections, the performance of only a few \amdl\ designs was compared. To get a complete comparison overview, Figure~\ref{gds_comparison1MeV} shows a comparative study of the performance for the effective area, angular/energy resolution, and sensitivity of all the five modules (from Figure~\ref{payload_design}) with their respective crystal thicknesses ($d = 1\ \si{cm}$ and $d=1.5\ \si{cm}$), at 1 MeV. From the figure can it be noticed that a \say{pair} of designs exists where the achieved performance is similar. One pair is formed by \mgdsthick\ and \mstack\ (designs \#2 and \#3), and another pair is formed by \mstackthick\ and \mstackthree\ (designs \#4 and \#5). For example, the performance achieved by \mgdsthick\ is slightly better to \mstack\ in terms of ARM and sensitivity, and so is the performance of \mstackthick\ slightly better to \mstackthree\ (except for angular resolution in that case). Therefore it can be concluded that incorporating \underline{one $d=1.5$ cm} crystal layer is more efficient than incorporating \underline{two $d=1$ cm} crystal layers, etc. Overall the \mstackthick-\mstackthree\ pair performs best for all parameters (energy resolution excluded). At the time of setting up the detector, several constraints (e.g., financial budget, payload size, electronics, power consumption, etc.) may occur. The purpose of this comparison study and Figure~\ref{gds_comparison1MeV} is to provide the payload scientist a leverage in judgement while selecting a suitable design based on the constraints mentioned above. 

\begin{figure*}[ht]
    \centering \includegraphics[width=0.9\textwidth]{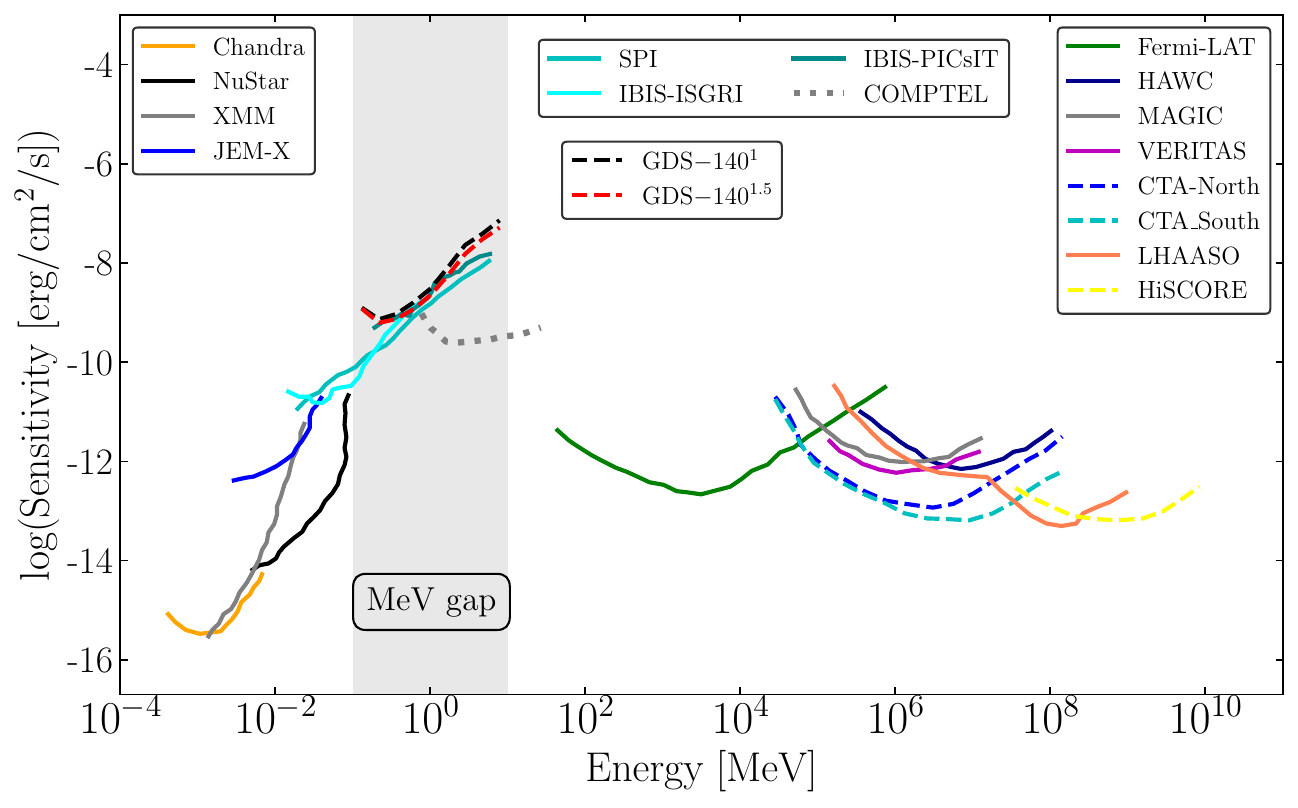}
    \caption{Comparison among continuum sensitivities of active instruments with \mgds and \mgdsthick. Other instruments' continuum sensitivities are adopted from \cite{2022JCAP...08..013L}. Past/current/future missions are shown in dotted/solid/dashed lines respectively. The MeV gap is shown as a grey band.}
    \label{sensi}
\end{figure*}

Finally, a comparison of \mgds\ and \mgdsthick\ sensitivity with other instruments is shown in Figure~\ref{sensi}. Here, the active missions are plotted with a solid line. Past missions (e.g., COMPTEL) are marked with dotted lines and future missions are plotted with dashed lines. 
When comparing the sensitivity to other instruments such as IBIS and COMPTEL in Figure \ref{sensi}, we note that the achieved sensitivity for both \mgds\ and \mgdsthick\ comes close to the sensitivity of IBIS up from 0.2 to 2 MeV. The standard \amdl\ designs could, therefore serve as a replacement for IBIS in this energy range. However, when \mgdsthick\ is extended to a 3U configuration,
it can reach comparable sensitivity to COMPTEL below 0.9 MeV, and even achieves better/comparable sensitivity to IBIS and SPI across most of the MeV gap. The achieved sensitivity for the 3U configuration is not as good as the untracked case for the payload design described in \cite{diwan20236u}, 
but this 3U payload (based on the 1U designs studied in this paper) incorporates a much simpler and cheaper design, and allows more available space for other instruments within the 3U volume. Nonetheless, even the standard 1U designs, \mgds and \mgdsthick, may be of scientific interest given their near-IBIS sensitivity achieved with a simple detector design.

\section{Conclusion}
Based on the simulations, different configurations of CZT payload options with various numbers of IDEAS MeV CZT crystal modules are compared.  
In this study, we also find that the effective area, ARM, and sensitivity are directly proportional to the payload size and crystal thickness. However, the energy resolution does not show any such clear correlation with the payload size.  From this comparative simulation study, constraints on electronics and cost give the edge in the procurement of the commercial off-the-shelf (COTS) components for the payloads. 
Payloads that are confined within the dimensions of a CubeSat and employ CZT technology for detecting $\gamma$-rays possess the capability to make substantial contributions towards the understanding of the suitability of modern-day technology for satellite missions that concentrate on the MeV energy range. Irrespective of the final choice of $\gamma$-ray detection technology, data derived from CubeSats can help in the calibration of forthcoming missions. Additionally, MeV payloads that utilise CubeSat technologies can serve as cost-effective pathfinders for future space technologies and missions. 
The laboratory performance and calibration testing of the new CZT detector design will be reported in a future publication.

\backmatter
\bmhead{Acknowledgements}
This research is supported by HKU-RMGS Funds (207300301,207301033; P.I.: Prof.~Q.~A.~Parker).
The research of P.~S.~Pal is partially supported by a General Research Fund (GRF) grant from the Research Grants Council of the Hong Kong Special Administrative Region, China (HKU Project 17304920; P.I. Dr.~Stephen C.~Y.~Ng.
We acknowledge IDEAS and Redlen Technologies for their technical support for simulation.)

\bibliography{reference}

\newpage 

\end{document}